\documentclass[journal,10pt]{IEEEtran}
\usepackage{algpseudocode}
\usepackage{algorithm}
\usepackage{algpseudocode}
\usepackage{amsmath}
\usepackage{multirow}
\usepackage{graphicx}
\usepackage{caption}
\usepackage{booktabs}
\usepackage{multirow}
\usepackage{url}
\usepackage{amsfonts}
\usepackage{amssymb}
\usepackage{graphicx} %use graph format
\usepackage{subfigure}
\usepackage{caption}
\usepackage{color}

\usepackage{cite}
\begin{document}

\title{Building Transmission Backbone for Highway Vehicular Networks: Framework and Analysis}

\author{Changle~Li,~\IEEEmembership{Senior Member,~IEEE}, Yao~Zhang,
Tom H. Luan,~\IEEEmembership{Member,~IEEE}, and Yuchuan~Fu

\thanks{Copyright (c) 2015 IEEE. Personal use of this material is permitted. However, permission to use this material for any other purposes must be obtained from the IEEE by sending a request to pubs-permissions@ieee.org. \emph{(Corresponding author: Changle Li.)}}
%\thanks{Manuscript received December 25, 2017. This work was supported by the National Natural Science Foundation of China under Grant No. 61571350, No. %61401334 and No. 61601344.}
\thanks{This work was supported by the National Natural Science Foundation of China under Grant No. 61571350 and No. 61601344, Key Research and Development Program of Shaanxi (Contract No. 2017KW-004, 2017ZDXM-GY-022), and the 111 Project (B08038).}
\thanks{C. Li, Y. Zhang, T. H. Luan, and Y Fu are with the State Key Laboratory of Integrated Services Networks, Xidian University Xi'an 710071, China (clli@mail.xidian.edu.cn, yzhang${\_}$01@stu.xidian.edu.cn, tom.luan@xidian.edu.cn, ycfu@stu.xidian.edu.cn).}% <-this % stops a space
}

%\markboth{IEEE Transactions on Vehicular Technology,~Vol.~XX, No.~XX, XXX~2017}
{}
%{Shell \MakeLowercase{\textit{et al.}}: Bare Demo of IEEEtran.cls for Journals}
\maketitle

\begin{abstract}

The highway vehicular ad hoc networks, where vehicles are wirelessly inter-connected, rely on the multi-hop transmissions for end-to-end communications. This, however, is severely challenged by the unreliable wireless connections, signal attenuation and channel contentions in the dynamic vehicular environment. To overcome the network dynamics, selecting appropriate relays for end-to-end connections is important. Different from the previous efforts (\emph{e.g.}, clustering and cooperative downloading), this paper explores the existence of stable vehicles and propose building a stable multi-hop transmission backbone network in the highway vehicular ad hoc network.
Our work is composed of three parts. Firstly, by analyzing the real-world vehicle traffic traces, we observe that the large-size vehicles, \emph{e.g.}, trucks, are typically stable with low variations of mobility and stable channel condition of low signal attenuation; this makes their inter-connections stable in both connection time and transmission rate. Secondly, by exploring the stable vehicles, we propose a distributed protocol to build a multi-hop backbone link for end-to-end transmissions, accordingly forming a two-tier network architecture in highway vehicular ad hoc networks. Lastly, to show the resulting data performance, we develop a queueing analysis model to evaluate the end-to-end transmission delay and throughput.

Using extensive simulations, we show that the proposed transmission backbone can significantly improve the reliability of multi-hop data transmissions with higher throughput, less transmission interruptions and end-to-end delay.

\end{abstract}


\IEEEpeerreviewmaketitle

\section{Introduction}\label{SectionI}

Highway travels, especially those on rural interstate highways, are important in our daily lives. As indicated in \cite{website}, one third of vehicle-miles driven in U.S. are on rural roads. In this case, connecting vehicles on highways as an integrated communication network can bring a variety of novel and exciting applications to the travelers. For example, \cite{Felice2014A} develops a distributed video streaming protocol to transmit live video streams to vehicles over the multi-hop inter-vehicular connections. As a result, vehicles can see real-time traffic video reports captured and transmitted from vehicles in front. \cite{Wang2013Dynamic} develops a multi-hop transmission protocol to enable popular content distribution, \emph{e.g.}, news, local advertisements, to vehicles over highway vehicular ad hoc networks so that vehicles can enjoy rich data information while on the move. \cite{Luan2015Social} proposes a vehicular social network on highways over which vehicles in proximity self-organize into a social network and share the mutually interested trip-related information using the pure vehicle-to-vehicle (V2V) connections. \cite{zhou2014chaincluster, wang2011delay, wang2012multicast, chen2017throughput} develop the cooperative schemes to provide broadband Internet services or reliable multicast traffic to vehicles using the combination of V2V and V2I communications.

%\cite{zhou2014chaincluster}, \cite{Chen2017Throughput}, \cite{wang2011delay} and \cite{wang2012multicast} develop cooperative schemes to provide broadband Internet services or reliable multicast traffic to vehicles using the combination of V2V and vehicle-to-roadside infrastructure communications.

The examples above all rely on the efficient multi-hop data transmissions over the ad hoc connected vehicles. This, however, is challenging in three aspects. 1) \emph{Unreliable connections}: With diverse velocities of vehicles, V2V connections are intermittent and unreliable, making the multi-hop transmissions susceptible to frequent interruptions. Measurement results reported in \cite{Zahn2009Feasibility} show that the average inter-vehicle connection time on highways is only about $15$ seconds. As a result, the efficiency of multi-hop data transmissions on V2V connected vehicles cannot be ensured. 2) \emph{Signal block}: The vehicles on highways may be obstacles to each other, resulting in significant signal attenuation. Specifically, the vehicles on highways are quite heterogenous in shapes and sizes; the large vehicles, \emph{e.g.}, trucks, can be moving obstructions to the small vehicles, \emph{e.g.}, compact cars, and block the wireless signals among small vehicles. For example, \cite{boban2011impact} shows that the additional signal attenuation due to large vehicles is about eight times higher than that in free space. 3) \emph{Channel contentions}: The highway vehicular networks are typically large-scale with a good many vehicles contending for transmissions in a small range. For example, within the communication range of $300$ meters that is assumed as the maximum communication range of vehicles in our study, there are about $20$-$35$ vehicles under smooth traffic flow and $206$-$283$ vehicles under jam traffic in a six-lane bidirectional highway \cite{May1990Traffic, Ren2016An}.

\begin{figure*}[!ht]
\centering
\subfigure[Standard deviation of speed in US-$101$]{
\label{1std.deviation}
\includegraphics[width=3.6in,height=2.4in]{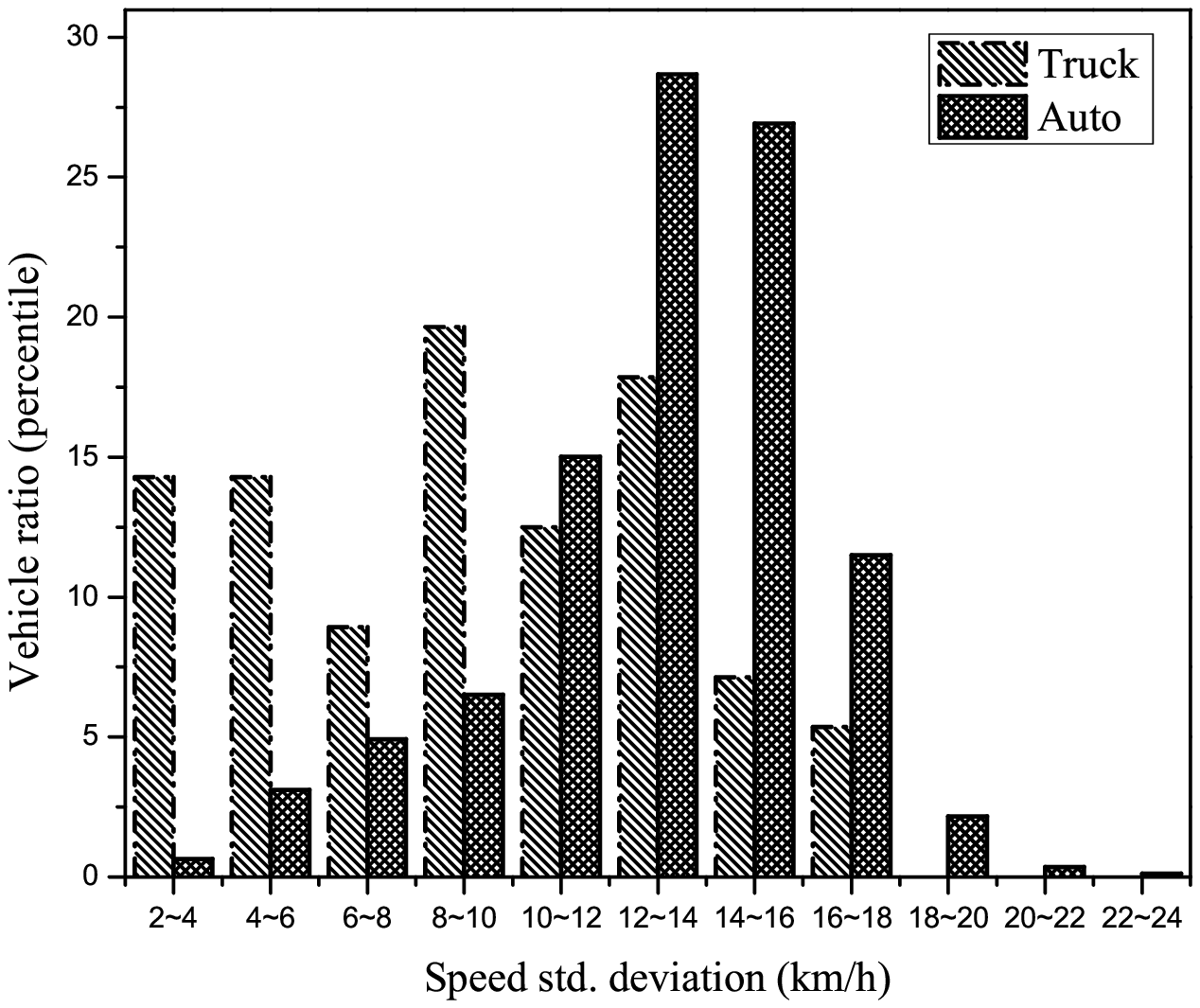}
}\hspace{-12ex}
\subfigure[Standard deviation of speed in I-$80$]{
\label{2std.deviation}
\includegraphics[width=3.6in,height=2.4in]{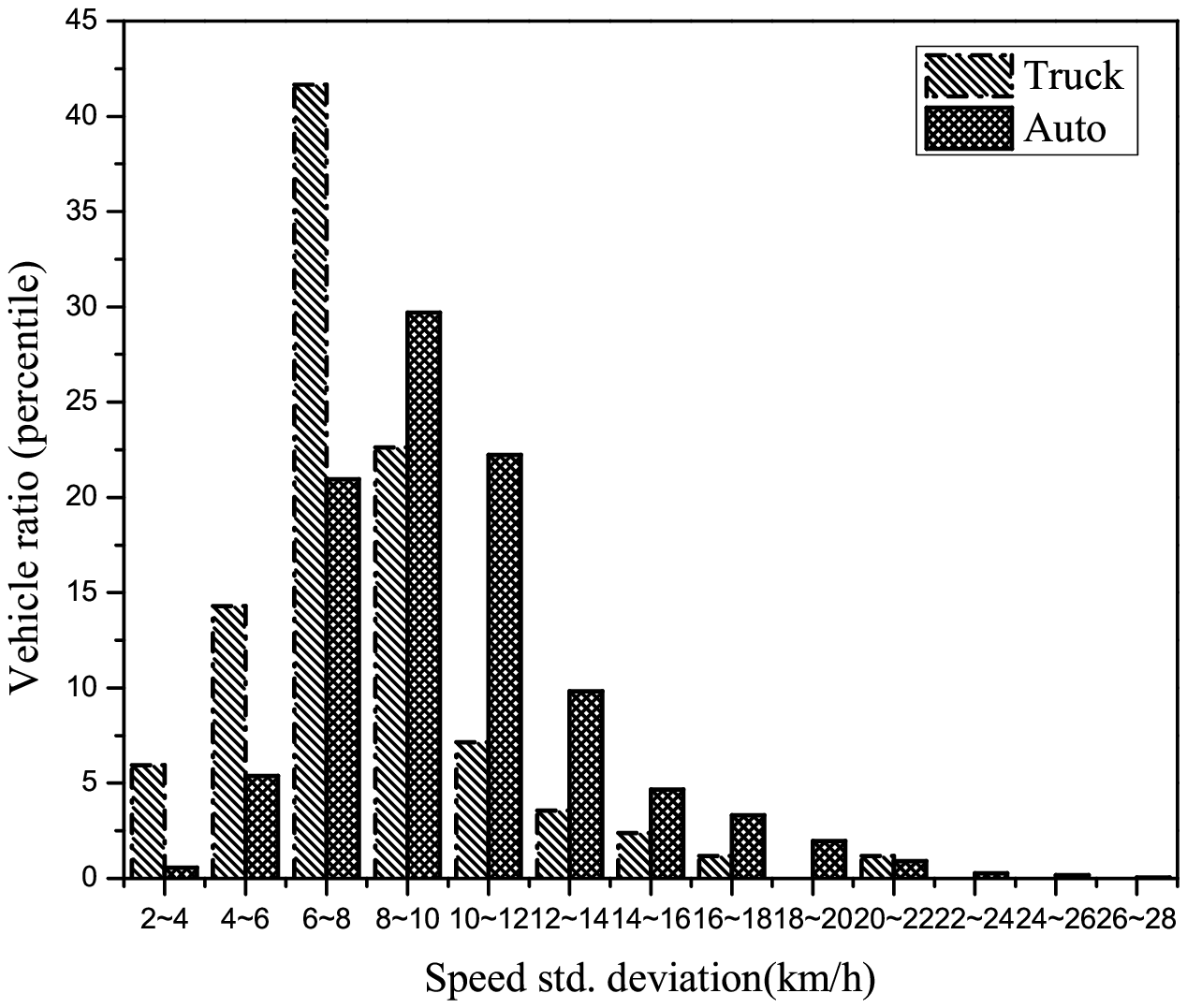}
}
 \caption{Results about standard deviation of speed}
 \label{Standard deviation}
\end{figure*}

The paper aims to enable efficient multi-hop data transmissions in highway vehicular ad hoc networks. Different from existing backbone based transmission schemes (\emph{e.g.}, SCRP \cite{togou2016scrp}), we explore the existence of stable vehicles in the network that have relatively stable inter-connection time and channel conditions, and then connect the stable vehicles as a transmission backbone. We proceed in three steps. 1) \emph{Framework Design}: To address the dynamic nature of vehicular networks, we study the highway vehicle traffics and propose a two-tier network architecture by exploring the stable vehicles in the highway. Specifically, based on the analysis of two highway traffic traces, we show that there exist a number of stable vehicles that have relatively stable communication capability as compared to others in terms of the inter-connection time and channel conditions. As such, we develop a two-tier vehicular network architecture as shown in Fig.~\ref{linkmodel}, where the top tier is built by connecting stable vehicles in a linear topology constituting a wireless transmission backbone for highway vehicular communications, and the bottom tier is formed in a star topology to cover other vehicles. To transmit data from one vehicle to another, data are first uploaded to stable vehicles and then transmitted over the top tier until reaching the destination. Note that the network is dynamic and the stable vehicles may depart from the backbone link, thus a distributed adaptive maintenance mechanism is adopted to maintain the network topology. 2) \emph{Analysis Model}: We develop an analytical model to evaluate the end-to-end performance of the proposal. Specifically, we model the vehicular network as connected queues, where each stable vehicle is represented by a $G/G/1$ queue. By applying the networked queueing analysis, we show the end-to-end transmission delay and achievable throughput of the network in closed expressions. 3) \emph{Verification}: Using extensive simulations, we verify the performance of our proposal and show that it outperforms the typical distributed multi-hop transmission scheme.
%\begin{figure}[!t]
%\centering
%\subfigure[Standard deviation of speed in US-101]{
%\begin{minipage}[b]{0.5\textwidth}
%\label{1std.deviation}
%\includegraphics[width=3.6in]{figure/1std.deviation.eps}
%\end{minipage}
%}
%\subfigure[Standard deviation of speed in I-80]{
%\begin{minipage}[b]{0.5\textwidth}
%\label{2std.deviation}
%\includegraphics[width=3.6in]{figure/2std.deviation.eps}
%\end{minipage}
%}
% \caption{Results about standard deviation of speed}
% \label{Standard deviation}
%\end{figure}

The remainder of this paper is structured as follows: Section \ref{SectionII} explores the existence of stable vehicles on highway and the model about signal attenuation. Section \ref{SectionIII} illustrates the system model including channel model and MAC layer model. Section \ref{SectionIV} has detailed introduction about the organization of the two-tier vehicular network architecture while the analysis framework is presented in Section \ref{SectionV}. Section \ref{SectionVI} evaluates the performance of our proposal. Finally, the related work and conclusion of our work are discussed in Section \ref{SectionVII} and Section \ref{SectionVIII}, respectively.
%\hfill August 4, 2015

\begin{figure*}[!ht]
\centering
\subfigure[Average speed in US-$101$]{
\label{1speed distribution}
\includegraphics[width=3.6in,height=2.4in]{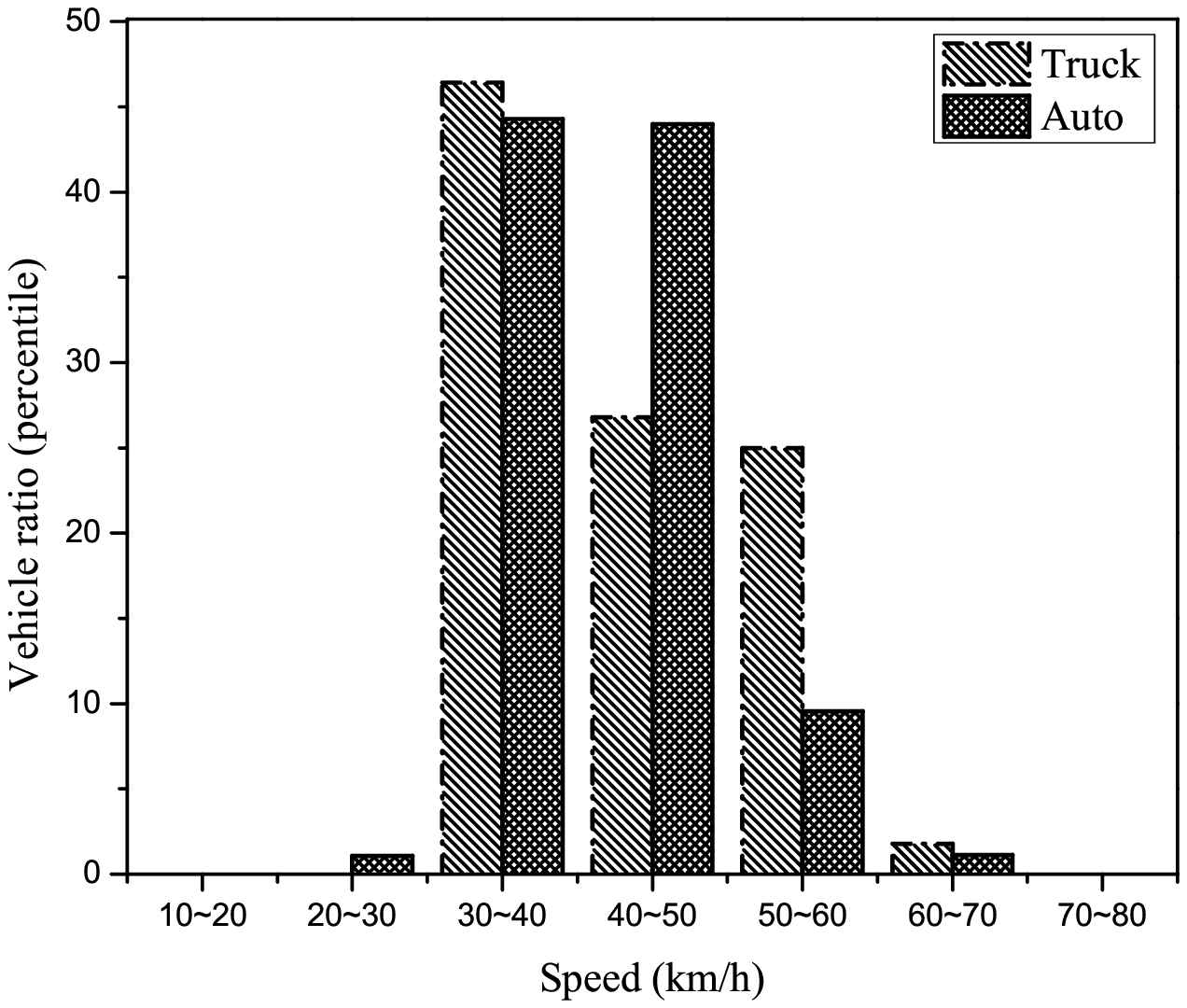}
}\hspace{-12ex}
\subfigure[Average speed in I-$80$]{
\label{2speed distribution}
\includegraphics[width=3.6in,height=2.4in]{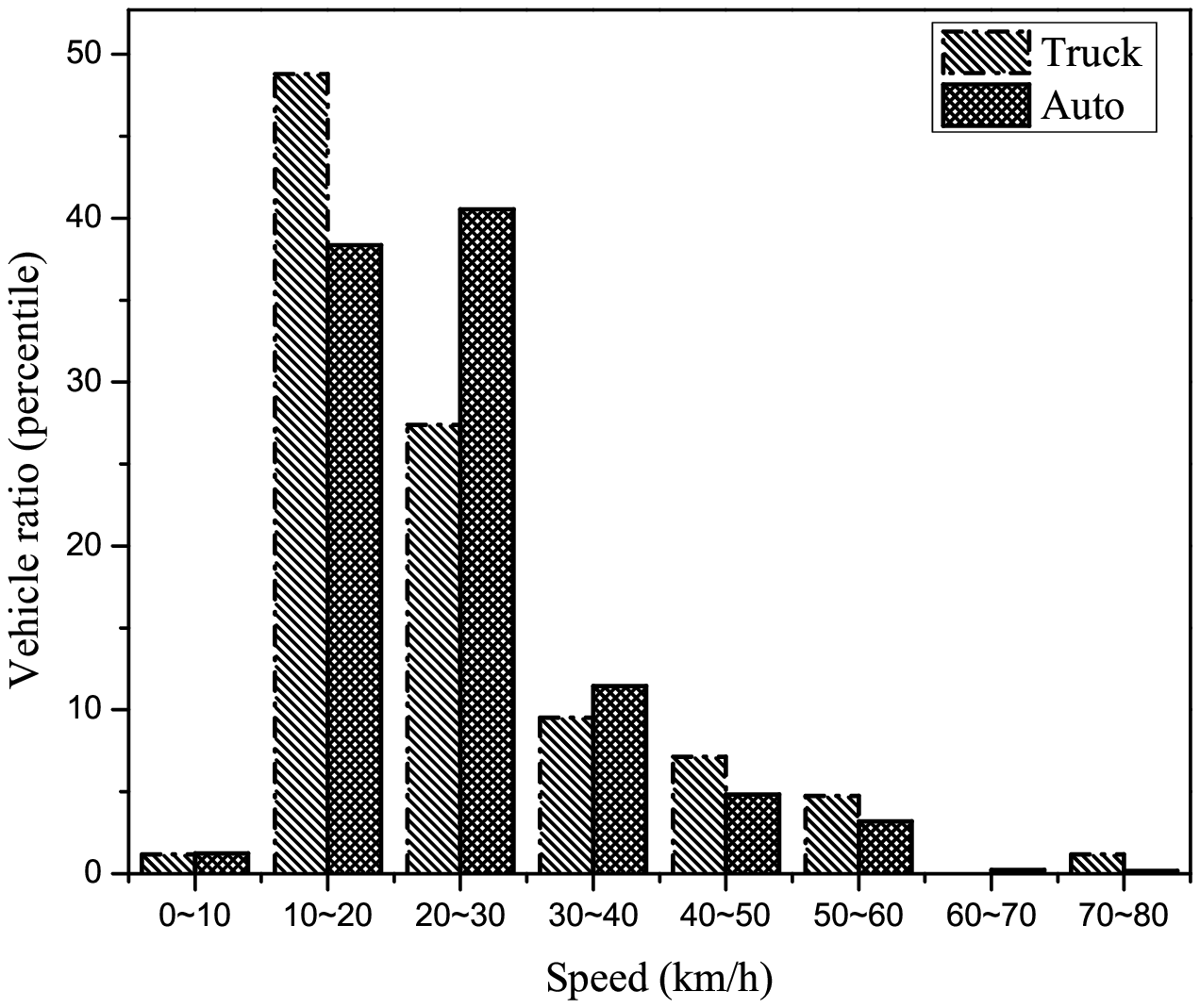}
}
 \caption{Results about average speed}
 \label{speed distribution}
\end{figure*}\vspace{-2ex}

\begin{figure*}[!ht]
\centering
\subfigure[Truck distribution in US-$101$]{
\label{truckratio_us101}
\includegraphics[width=3.0in,height=2.2in]{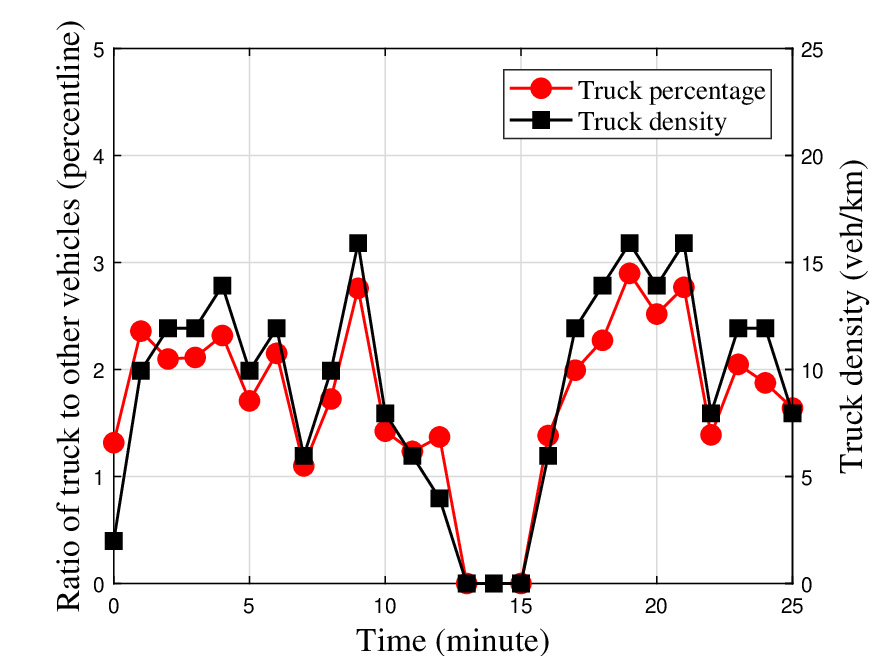}
}\hspace{-0ex}
\subfigure[Truck distribution in I-$80$]{
\label{truckratio_I80}
\includegraphics[width=3.0in,height=2.2in]{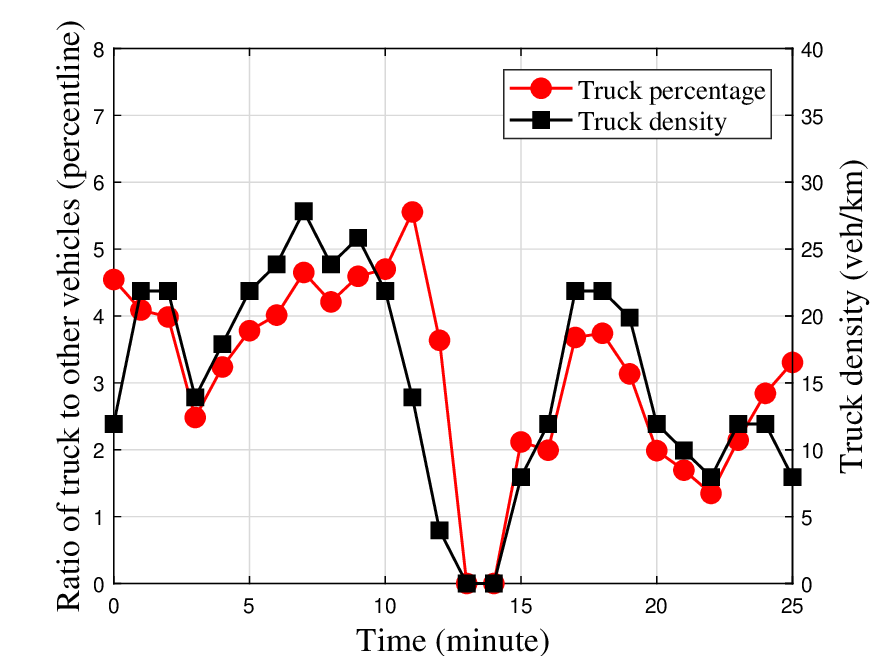}
}
 \caption{Results about truck distribution}
 \label{truckratio}
\end{figure*}

\section{Stable Vehicles: Existence and Model}\label{SectionII}

%\begin{figure}[!t]
%\centering
%\subfigure[Average speed in US-101]{
%\begin{minipage}[b]{0.5\textwidth}
%\label{1speed distribution}
%\includegraphics[width=3.7in]{figure/1averagespeedofvehicle_us101.eps}
%\end{minipage}
%}
%\subfigure[Average speed in I-80]{
%\begin{minipage}[b]{0.5\textwidth}
%\label{2speed distribution}
%\includegraphics[width=3.7in]{figure/2averagespeedofvehicle_i80.eps}
%\end{minipage}
%}
% \caption{Results about average speed}
% \label{speed distribution}
%\end{figure}

Our study relies on highway stable vehicles ($SV$s) to form backbone connections for data delivery. The $SV$s are defined in two aspects, \emph{i.e.}, stable inter-connection time and stable channel conditions. In this section, we first identify the existence of $SV$s by analyzing the real-world vehicular trajectory data collected from highway US-$101$ and I-$80$, and then model them.

%\begin{figure*}[!tb]
%  \centering
%  \begin{minipage}[t]{0.29\textwidth}
%    \includegraphics[width=2.6in,height=1.8in]{figure/1std.deviation.eps}
%\caption{Standard deviation of speed in US-101}
%\label{1std.deviation}
%  \end{minipage}
%  \hspace{-0.2ex}
%  \begin{minipage}[t]{0.29\textwidth}
%    \includegraphics[width=2.6in,height=1.8in]{figure/2std.deviation.eps}
%\caption{Standard deviation of speed in I-80}
%\label{2std.deviation}
%  \end{minipage}
%  \hspace{-0.2ex}
%  \begin{minipage}[t]{0.29\textwidth}
%  \includegraphics[width=2.6in,height=1.8in]{figure/1averagespeedofvehicle_us101.eps}
%\caption{Average speed in US-101}
%\label{1speed distribution}
%  \end{minipage}
%  \hspace{2.5ex}
%\end{figure*}
%
%\begin{figure*}[!tb]
%  \centering
%  \begin{minipage}[t]{0.29\textwidth}
%    \includegraphics[width=2.6in,height=1.8in]{figure/2averagespeedofvehicle_i80.eps}
%\caption{Average speed in I-80}
%\label{2speed distribution}
%  \end{minipage}
%  \hspace{2.3ex}
%  \begin{minipage}[t]{0.29\textwidth}
%    \includegraphics[width=2.2in,height=1.65in]{figure/truckratio_us101.eps}
%\caption{Truck distribution in US-101}
%\label{truckratio_us101}
%  \end{minipage}
%  \hspace{0.1ex}
%  \begin{minipage}[t]{0.29\textwidth}
%  \includegraphics[width=2.2in,height=1.65in]{figure/truckratio_i80.eps}
%\caption{Truck distribution in I-80}
%\label{truckratio_I80}
%  \end{minipage}
%  \hspace{3ex}
%\end{figure*}

\subsection{Velocities of Highway Vehicles}

%\begin{figure}[!t]
%\centering
%\subfigure[Truck distribution in US-101]{
%\begin{minipage}[b]{0.5\textwidth}
%\label{truckratio_us101}
%\includegraphics[width=3.2in]{figure/truckratio_us101.eps}
%\end{minipage}
%}
%\subfigure[Truck distribution in I-80]{
%\begin{minipage}[b]{0.5\textwidth}
%\label{truckratio_I80}
%\includegraphics[width=3.2in]{figure/truckratio_i80.eps}
%\end{minipage}
%}
% \caption{Results about truck distribution}
% \label{truckratio}
%\end{figure}

We first verify the existence of the small portion of $SV$s in highway scenarios according to a series of analysis results as shown in Fig \ref{Standard deviation}-\ref{truckratio}. Specifically, we study the vehicle trajectory of Next Generation Simulation (NGSIM) data \cite{NGSIM}, which is collected from U.S Highway $101$ (US-$101$) in Los Angeles, California. In this data set, we select a total of $25$ minutes (from $8:05$ a.m. to $8:30$ a.m.) of data in the morning time from a $640$ meters length of the study area. For comparison, we conduct the same analysis on the I-$80$ vehicle data set collected from Interstate $80$ (I-$80$) in Emeryville, California. The length of study area is $503$ meters and the time period is also $25$ minutes (from $5:05$ p.m. to $5:30$ p.m.) in the afternoon time. The total numbers of vehicles surveyed during the study period are $3917$ (US-$101$) and $3344$ (I-$80$), respectively. The observation results from different data sets at different time periods and locations can improve the generality of obtained conclusion.

Intuitively, the mobility characteristics of vehicles can be reflected by their speed fluctuation. The conventional view is that moving vehicles are by nature dynamic so that they cause intermittent connections and dynamic networks. However, what we focus on in this paper is the stability of vehicles. Let us assume that $SV$s are the vehicles that have low rate of speed change and low relative mobility to their neighbors. The data sets contain three types of vehicles (truck, automobile and motorcycle) and present the vehicle trajectory in highways. For simplicity, we select truck data and auto data to represent the large vehicles (such as truck, van and bus) and general vehicles (such as private vehicles) on the road, respectively. Firstly, we compute the standard deviation of speed for all the vehicles in two highways and present the statistical distribution in Fig.~\ref{Standard deviation}. In the figures, most of trucks have the lower standard deviations than autos. For example, in the vehicle trajectory of US-$101$, we note that about $69.6\%$ of trucks have the standard deviations of less than $12$ while this ratio among autos is only $30.2\%$. Similarly, in the vehicle trajectory of I-$80$, the ratio of trucks that have the standard deviations of less than $10$ is about $84.5\%$ while this ratio among autos is about $56.6\%$. In conclusion, trucks have relatively stable speeds whereas the autos are quite dynamic with the larger variation in speeds.

In Fig.~\ref{1speed distribution} and Fig.~\ref{2speed distribution}, we plot
the speed distributions of trucks and autos in highway US-$101$ and I-$80$,
respectively. From Fig. \ref{1speed distribution}, it can be seen that the average speeds of trucks in different time have a major centralized range ($20-30$ km/h) which contains about $46.4\%$ trucks. By comparison, the average speeds of autos in different time have two major ranges ($20-30$ km/h and $30-40$ km/h), which account for $44\%$ and $43\%$, respectively. The same conclusion can be concluded in Fig. \ref{2speed distribution}, about $49\%$ trucks have the speed range from $10-20$ km/h while the two major speed ranges of autos ($10-20$ km/h and $20-30$ km/h) account for $38\%$ and $43\%$, respectively. Compared with autos, trucks have more centralized speed distribution which means there is low relative mobility among trucks. From above analysis on speed standard deviation and relative mobility, we note that there are a part of vehicles that have relative stability in highways.

To utilize the small portion of vehicles in vehicular networks, we further exploit the changes of truck percentage and density over time, as shown in Fig.~\ref{truckratio_us101} and Fig.~\ref{truckratio_I80}. The reason that
both figures have several breakpoints is that the trajectory samples in those
few minutes are not available. From Fig.~\ref{truckratio_us101}, it is clear
to see that the ratio of trucks to all vehicles is approximately $2\%$ to
$3\%$ during our study time. Meanwhile, the truck density is approximately $6$
veh/km ($166.7$ meters between vehicles) to $15$ veh/km ($66.7$ meters between
vehicles). As such, the density is sufficient to support
the stable connection between any two adjacent trucks when the maximum communication range of each vehicle is $300$ meters. Similarly, in the data set of I-80, the ratio of trucks
to all vehicles is about $2\%$ to $5\%$ while the truck density is about $8$
veh/km to $25$ veh/km, which can reaffirm our observation.

%\begin{figure*}[!htb]
%  \centering
%  \begin{minipage}[t]{0.3\linewidth}
%    \includegraphics[width=2.8in,height=2in]{figure/2averagespeedofvehicle_i80.eps}
%\caption{Average speed in I-80}
%\label{2speed distribution}
%  \end{minipage}
%  \hspace{2.8ex}
%  \begin{minipage}[t]{0.3\linewidth}
%    \includegraphics[width=2.2in,height=1.7in]{figure/truckratio_us101.eps}
%\caption{Truck distribution in US-101}
%\label{truckratio_us101}
%  \end{minipage}
%  \hspace{1.5ex}
%  \begin{minipage}[t]{0.3\linewidth}
%  \includegraphics[width=2.2in,height=1.7in]{figure/truckratio_i80.eps}
%\caption{Truck distribution in I-80}
%\label{truckratio_I80}
%  \end{minipage}
%  \hspace{5ex}
%\end{figure*}

\subsection{Channel Conditions of Vehicles}

%When the obstructed area researches 60\% of the radius of first Fresnel zone, the signal attenuation on a radio link will increase significantly, which is independent of the number and position of obstructions \cite{boban2011impact}.

Due to heterogeneous shapes and sizes, vehicles may be moving obstacles to
each other and block the wireless signal. As reported in \cite{boban2011impact, He2014Vehicle} the moving vehicles will induce significant attenuation in the line of sight (LOS) between Tx and Rx.

\begin{figure}[!tb]
\centering
\includegraphics[width=3.4in]{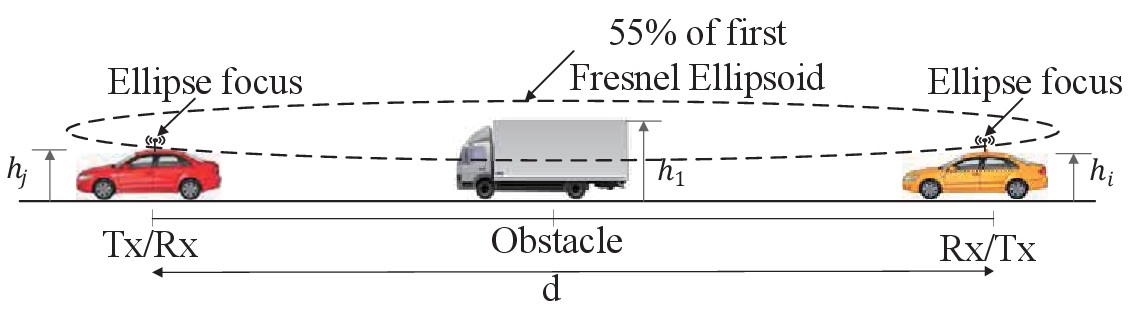}
\caption{Analysis scenario for the impact of the obstacle on LOS}
\label{Analysisscenario1}
\end{figure}

\begin{figure}[!tb]
\centering
\includegraphics[width=3.0in]{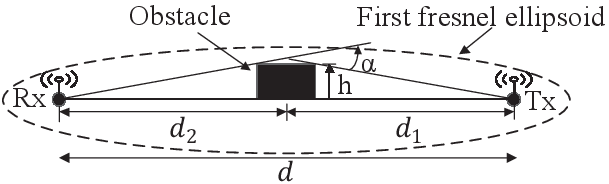}
\caption{Abstracted model for analysis scenario}
\label{abstractedfigure}
\end{figure}

Based on existing works, we further examine the impact of obstacles on V2V signal propagation and use the results to model the channel conditions of different types of vehicles. To this end, we employ a single knife-edge diffraction model to evaluate the signal attenuation of the V2V link obstructed by vehicles. By assuming the obstacle as a semi-infinite perfectly absorbing plane, the knife-edge diffraction model theoretically presents an adequate approximation for the diffraction of the electromagnetic waves \cite{Rappaport2002Wireless}. Since our aim is to illustrate the difference of two types of vehicles as obstacles, i.e., large vehicles (\emph{e.g.}, trucks) and small vehicles (\emph{e.g.}, private cars), we consider a simple scenario that there is one obstacle between Tx and Rx, as shown in Fig.~\ref{Analysisscenario1}. Fig. \ref{abstractedfigure} is the abstracted model.

Consider that the frequency of DSRC is $5.9$ GHz, we introduce a
Fresnel-Kirchoff diffraction parameter \emph{$v$} \cite{Rappaport2002Wireless}%
:
\begin{equation}
\label{Fresnel-Kirchoff1}v=h\sqrt{\frac{2(d_{1}+d_{2})}{\lambda{d_{1}}d_{2}}%
}=\alpha\sqrt{\frac{2{d_{1}}{d_{2}}}{\lambda({d_{1}}+{d_{2}})}}.
\end{equation}
The results of normal fit about the heights of large vehicles and small
vehicles are shown in Table~\ref{Table1}, by referring to
\cite{boban2011impact}. The additional signal attenuation $G_{d}(dB)$ caused
by two types of obstacles can be obtained by the following equation
\cite{propagationbydiffraction}:

\begin{table}[!t]
\renewcommand{\arraystretch}{1}
\newcommand{\tabincell}[2]{\begin{tabular}{@{}#1@{}}#2\end{tabular}}
\caption{Normal Fit for the Height of Vehicles}
\label{Table1}
\centering
\begin{tabular}{|c|c|c|c|}% |��ʾ��������
\hline%�±�����
\tabincell{c}{}&{Large vehicles} & \tabincell{c}{Personal vehicles} \\
\hline
mean (m) & 3.35 & 1.5\\
std.deviation (m) & 0.084 & 0.084\\
\hline
\end{tabular}
\end{table}

\begin{equation}
\label{signalattenuation}G_{d}(dB)=
\begin{cases}
6.9+ & \\
20log10[\sqrt{(v-0.1)^{2}+1}+v-0.1], & \mbox{v$>$ -0.7}\\
0, & \mbox{v$\leq$ -0.7}.
\end{cases}
\end{equation}

\begin{figure}[!tb]
\centering
\includegraphics[width=3.1in]{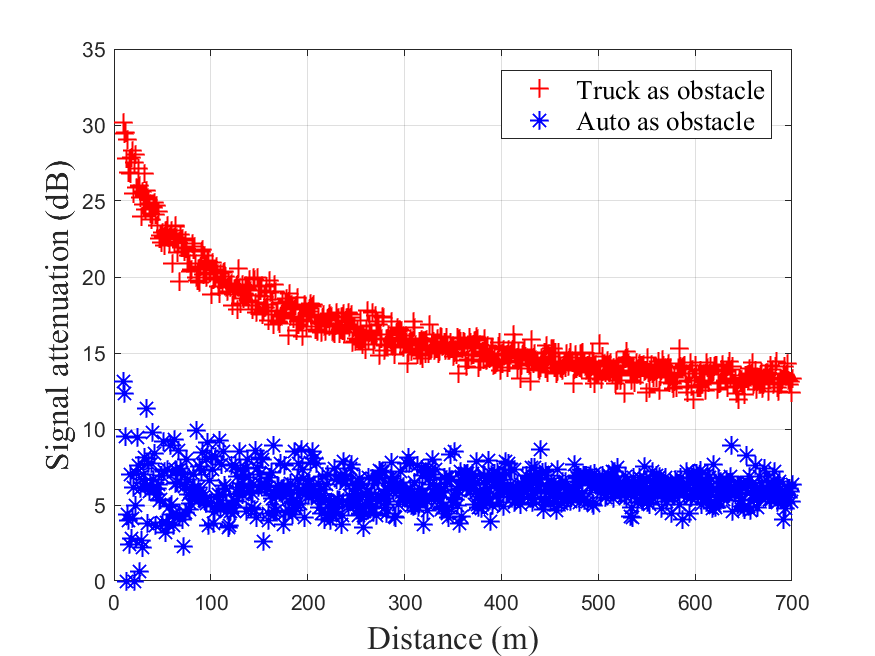}
\caption{Impact of truck and auto on signal propagation}
\label{signalattenution}
\end{figure}

Fig.~\ref{signalattenution} shows the analysis results of the diffraction gain. As reported in \cite{boban2011impact}, the knife-edge model is too optimistic to well characterize the signal attenuation at shorter distances, so that we adopt $10$ m as the starting inter-vehicle distance. We can see that there is a marked distinction between the impact of trucks and autos as obstacles on the signal propagation. The average additional attenuation caused by autos is about $6$ dB, while the value obtained by trucks is about $16$ dB which is corresponding to the analysis results ($15-20$ dB) about the bus obstruction in \cite{He2014Vehicle}. %{\color{red}{By averaging all types of vehicles, the signal attenuation is about $9.2$ dB according to \cite{boban2011impact} where the multiple knife-edge diffraction model is adopted.}}
From above analysis we can conclude that, large vehicles may incur more signal attenuation to smaller cars and accordingly result in more harsh wireless conditions for V2V communications. In other words, it shows that the wireless channel between large vehicles is more stable than that between small vehicles.

To conclude, by analyzing the mobility and channel conditions of large vehicles and small vehicles, we show that the large vehicles have both stable mobility and
channel conditions. The existence of $SV$s can be verified. In what follows, we present a distribute algorithm to form the data delivery path among vehicles by using $SV$s.

\section{System Model}\label{SectionIII}

In this section, we establish the system model to evaluate the transmission capacity of highway vehicular networks. To this end, we first model the wireless propagation channel, and then evaluate channel contentions by modeling the MAC layer.

\subsection{Channel Model}

We assume that all vehicles can communicate with each other within the communication range by equipping with on-board wireless devices. To
accurately identify the signal attenuation and transmission rate in the highway environment, we are ready to characterize the radio channel by modeling the large-scale path loss and small-scale fading. Typically, the case that buildings locate between communication pairs is not found in highway scenarios, so that we do not consider the non-LOS links blocked by buildings \cite{Nilsson2017A}. Based on above analysis about the impact of vehicle as obstacles, we divide the highway V2V radio link into two types: link-of-sight (LOS) and obstructed-line-of-sight (OLOS).
%In our study�����еij�������equipped with on-board transceivers ����ÿ������"each has a single antenna so that they can not transmit and receive at the same time"��In this section, we are ready to ����the radio link modeling the large-scale path loss and small-scale fading to characterize the wireless channel��

\subsubsection{Large-Scale Modeling}

In LOS links, there are no obstructions between communicating pairs. The
signal attenuation is mainly caused by path loss and large-scale fading. In
this case, we adopt the two-ray ground reflection model to characterize the
E-field \cite{Rappaport2002Wireless}:
\begin{equation}
\label{tworeflection}
    E(d,t)=\frac{E_{0}d_{0}}{d^{\prime}}cos(\omega_{c}(t-\frac{d^{\prime}}{c}))+(-1)\frac{E_{0}d_{0}}{d^{\prime\prime}}cos(\omega_{c}(t-\frac{d^{\prime\prime}}{c})),
\end{equation}
where $d^{\prime}$ is the direct LOS distance and $d^{\prime\prime}$ is the distance of ground-reflected path. Both of them are calculated based on the practical antenna heights of vehicles. $E_{0}$ is the known electric field strength at the reference distance $d_{0}$. The angular frequency can be
calculated as $\omega_{c}=2\pi/f$, where $f$ is the frequency of DSRC. $c$ is the light speed in free space.

%%�������϶Գ�����Ϊobstacle��Ӱ���ķ��������Ǹ����Ƿ��г��赲��V2Vlink��Ϊ���֣�line-of-sight (LOS) and obstructed-line-of-sight (OLOS).
%In LOS links��communication pairs֮��û���ϰ����赲����V2V link��Ӱ����Ҫ��path loss, large-scale fading.we Ӧ��the two-ray ground reflection model to characterize the E-field of LOS \cite{Rappaport2002Wireless}, as follows:
%\begin{equation}\label{tworeflection}
%E(d,t)=\frac{E_0d_0}{d'}cos(\omega_c(t-\frac{d'}{c}))+(-1)\frac{E_0d_0}{d''}cos(\omega_c(t-\frac{d''}{c})),
%\end{equation}
%where $d'$ is the direct LOS distance and $d''$ is the distance of ground-reflected path and both of them are calculated by the practical antenna heights. $E_0$ is the known electric field strength at the reference distance $d_0$. Besides, the angular frequency $\omega_c=2\pi/f$, where $f$ is the frequency of DSRC and $c$ is the light speed in free space.

Given the E-field at distance $d$, we can calculate the received power $P_{r}$
(in watts) as
\begin{equation}
\label{LOSpower}P_{r}=\frac{|E(d,t)|^{2}G_{r}\lambda^{2}}{480\pi^{2}},
\end{equation}
where $|E(d,t)|$ is the E-field envelope and $G_{r}$ is the antenna gain of
the receiver.

In OLOS links, we only consider the signal attenuation caused by vehicles. In highway radio links, it is estimated that there are two or more vehicles locating between communication pairs. As such, the multiple knife-edge model, which is the extension of the single knife-edge model, is applied to  model the propagation channel of OLOS links accurately \cite{Rappaport2002Wireless}. In this case, the Fresnel-Kirchoff diffraction parameter \emph{v} becomes
\begin{equation}
\label{Fresnel-Kirchoff2}v=h^{\prime}\sqrt{\frac{2(d_{1}+d_{2})}{\lambda
{d_{1}}d_{2}}},
\end{equation}
where $h^{\prime}$ is the equivalent height of multiple obstacles \cite{Rappaport2002Wireless}. By substituting (\ref{Fresnel-Kirchoff2}) into (\ref{signalattenuation}), the additional signal attenuation due to vehicles
can be obtained.
%to the height of knife-edge represented by the triangle included two antennas and top of the peak that blocks the line-of-sight from the antenna.
%Ȼ��ͨ����ʽEq.ref{signalattenuation}�������ɶ������������Ķ����ź����ġ�

\subsubsection{Small-scale modeling}

We implement the zero-mean normal model to evaluate the small-scale fading for
LOS and NLOS \cite{Boban2014Geometry}. For each radio link,
%For link $i$ ($i$=0 represents the LOS and $i$=1 represents the OLOS),
the small-scale signal deviation $\sigma_{i}$ is
\begin{equation}
\label{signaldeviation}
\sigma=\sigma_{\min}+\frac{\sigma_{\max}-\sigma_{\min}}{2}\left(\sqrt{\frac{NV}{NV_{\max}}}+\sqrt{\frac{AS}{AS_{\max}}} \right)  ,
\end{equation}
where $\sigma_{\min}$ and $\sigma_{\max}$ are the minimum and maximum deviation value, respectively. $NV$ is the number of vehicles per unit area in the communication range while $NV_{\max}$ is the maximum $NV$. $AS$ is the area of static obstructions per unit area and $AS_{\max}$ is the maximum $AS$. However, since we only consider moving obstruction, the $AS$ can be neglected.

Using the large-scale and small-scale model, we calculate the received power as%

\begin{equation}
\label{receivedpower}
P_{rTOT}=10log_{10}(P_{r})+N(0,\sigma).
\end{equation}

In order to determine the transmission rate of the wireless transceiver, we adopt
the channel modulation method defined in DSRC standard \cite{Astm2003Standard}%
. IEEE 802.11p physical layer supports four modulation schemes and three FEC
coding rates to compute eight transmission rates. The mapping table that shows
the relationship between minimum sensitivity thresholds and transmission rates
is shown in TABLE \ref{SNR} \cite{boban2011impact}.

\begin{table}[htb]
 \caption{Received Power Threshold and Data Rate}
 \label{SNR}
 \begin{tabular}{lclclclclclclcl}
  \toprule
  Threshold (dBm): &-85  &-84 &-82  &-80 &-77 &-70  &-69 &-67  \\
  \midrule
  Data Rate (Mbps):  & 3 & 4.5 & 6& 9 & 12 & 18 & 24 & 27\\
  \bottomrule
 \end{tabular}
\end{table}

%\begin{table}[htbp]
%\caption{\label{SNR}}
%\begin{tabular}{lclclclclclclcl}
%\toprule
%SNR Threshold (dB):  & 5 & 6 & 8& 11 & 15 & 20 & 25 &N/A\\
%\midrule
%Data Rate (Mbps): &3 &4.5 &6 &9 &12 &18 &24 &27 \\
%\bottomrule
%\end{tabular}
%\end{table}
%For example, if the SNR of received signal is 10 dB which is above the threshold 8 dB, the transmitter will select 6 Mbps as the transmission rate.

\subsection{MAC Layer Model}

In our study, we employs the contention-based MAC protocol to characterize the channel contentions among parallel transmissions. Since we focus on the efficiency of data transmissions without prioritization requirements, the IEEE 802.11b distributed coordination function (DCF) is applied for MAC scheduling.

Let $\tau_{p}$ denote the probability that vehicles transmit packets in a slot time.
We assume that \emph{CW} represents the \emph{Contention Window} size, which is used to compute the backoff time of vehicles. We have
\begin{equation}
\label{attemptprobility}\tau_{p}=\frac{2}{CW+1}.
\end{equation}

According to existing statistical analysis in \cite{Bai2009Spatio}, the highway vehicle traffic flow can be modeled as the exponential model. Therefore, the Poisson distribution is used to estimate the number of vehicles within the communication range. The probability that there are \emph{n} vehicles locating within the communication range \emph{s} is
\begin{equation}
\label{Poisson}P(V_{num}=n)=\frac{({\gamma}s)^{n}e^{(-{\gamma}s)}}{n!},
\end{equation}
where $\gamma$ is the road traffic density.

According to the transmission attempt probability $\tau_{p}$, the probability of no transmissions in a slot time is $(1-\tau_{p})^{n}$. So that the probability that at least one vehicle transmits data in the channel can be
obtained as
\begin{equation}
\label{onetransmission}
  P_{t}=1-(1-\tau_{p})^{n}.%
\end{equation}

We assume that $p_{e}$ is the error probability of transmissions on channel. We can obtain the successful transmission probability that one vehicle transmits packets in a randomly slot time is
\begin{equation}
\label{successfulprobability}
P_{s}=n\tau_{p}(1-\tau_{p})^{n-1}(1-p_{e}),
\end{equation}
where the first part means that $n-1$ remaining vehicles have no transmissions in the slot time.

By (\ref{onetransmission}) and (\ref{successfulprobability}), we can calculate the average length of a time slot in DCF as

\begin{equation}
\label{MeanLength}
T=(1-P_{t})\delta+(P_{t}-P_{s})T_{c}+P_{s}T_{s},
\end{equation}
where $\delta$ denotes $SlotTime$. $T_s$ and $T_c$ are the successful transmission time and collision time, respectively. According to \cite{zhou2014chaincluster, Bianchi2002Performance, luan2013integrity}, we have

\begin{equation}
\label{T_c}
T_c=RTS+DIFS+\delta,
\end{equation}

\begin{equation}
\label{T_s}
T_s=RTS+3SIFS+4\delta+CTS+ACK+DIFS+\frac{E(F)}{E(C)},
\end{equation}
where $E(F)$ and $E(C)$ are the average frame length and modulation rate, respectively.
%
%The throughput is defined as the fraction of transmitted payload size to
%transmission time. According to existing results about DCF
%\cite{zhou2014chaincluster}\cite{Bianchi2002Performance}, we can calculate the
%throughput of MAC layer as follows
%\begin{equation}
%\label{throughput}R_{thr}=\frac{P_{s}(E[P]+H)}{(1-P_{t})\delta+(P_{t}%
%-P_{s})T_{c}+P_{s}T_{s}},
%\end{equation}
%where $E[P]$ and $H$ is the payload and header length of transmitted packet,
%$\delta$ is the $SlotTime$, $T_{s}$ and $T_{c}$ are the successful
%transmission time and collision time, respectively. The detailed theoretical
%derivations of $T_{s}$ and $T_{c}$ can be obtained in \cite{luan2013integrity}
%while the difference is that we assume the frame length is constant.

\section{Multi-Hop Transmission using Tow-Tier Vehicular Network Architecture}\label{SectionIV}

\begin{figure}[!tb]
\centering
\includegraphics[width=3.4in,height=1.6in]{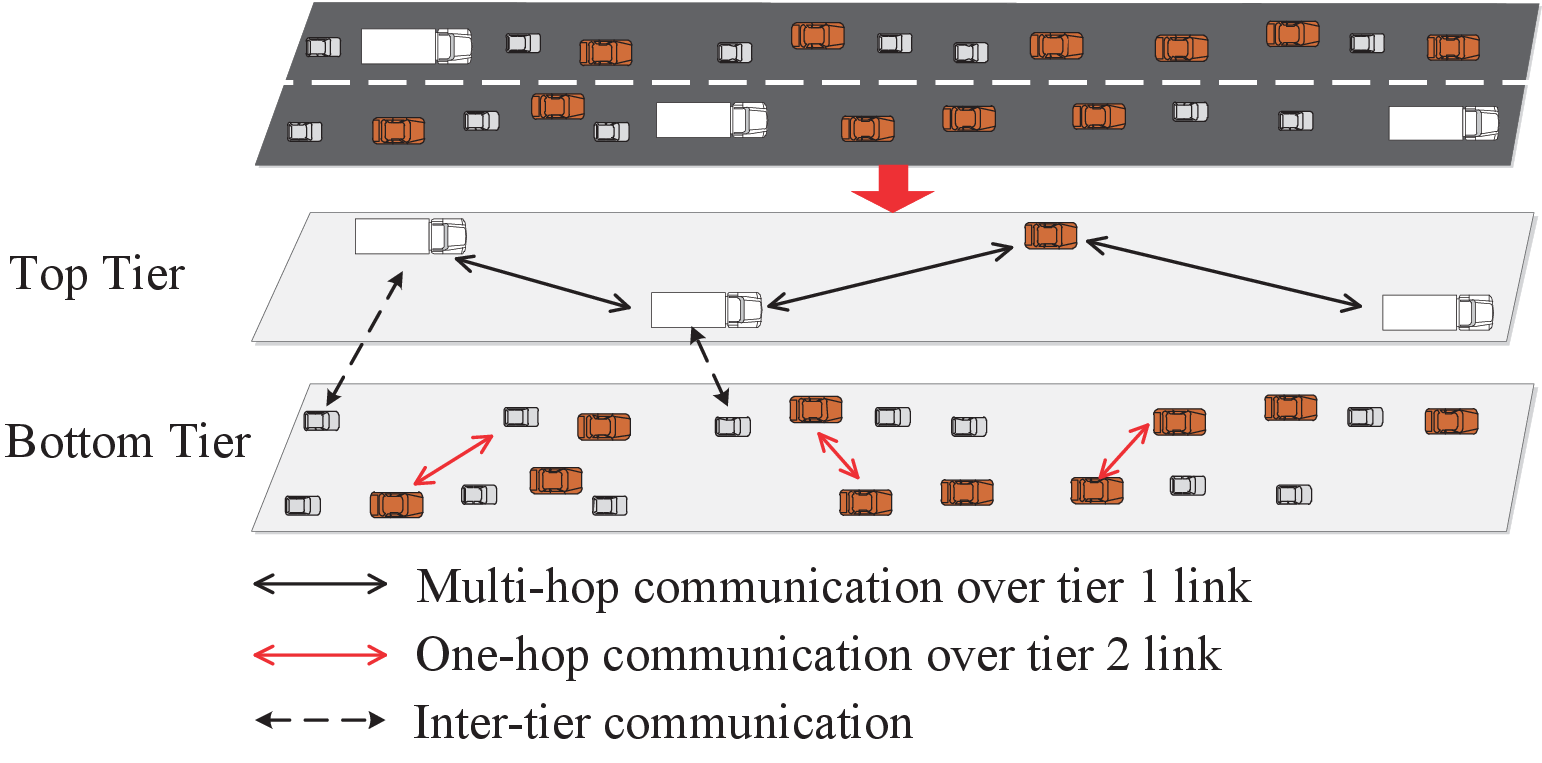}
\caption{Two-tier vehicular network architecture on highway}%
\label{linkmodel}%
\end{figure}

Above analysis on the existence and influence of $SV$s enables us to explore their roles in highway inter-vehicle communications. To optimize the utilization of these valuable vehicles, we propose a two-tier vehicular network architecture, as shown in Fig.~\ref{linkmodel}. The top tier is comprised of the $SV$s selected from vehicles while the bottom tier includes other vehicles that connect to the top tier nodes. Because of the relatively stable characteristics of members, the top tier has more reliable V2V links, making it suitable for multi-hop backbone transmissions. We should stress that although the conclusion of $SV$s is obtained based on two real-world vehicle traces, our newly proposed scheme of selecting $SV$s and the network architecture are general since they do not depend on any vehicle traces.

To reduce the interference among vehicles, it is assumed that each vehicle is equipped with two DSRC transceivers. Transceiver 1 is used to broadcast beacon messages while transceiver 2 is for data transmissions. Similar method is used in multiplatoon communications and real-world measurements to evaluate the network performance,\cite{Nilsson2017A},\cite{Peng2017Performance}. The practical solution can also be found from Kapsch TrafficCom AB \cite{website2} whose transceiver has the function of dual radio support for either concurrent or redundant channel operation. We also assume that all vehicles can obtain their own locations by the on-board GPS devices. Moreover, it is necessary to develop an adaptive maintenance mechanism since $SV$s may depart from the network. A brief introduction about the construction of a transmission backbone link is shown in Algorithm \ref{Algorithm1}. We next present the detailed organization of the two-tier network architecture.

\floatname{algorithm}{Algorithm}
 \begin{algorithm}
        \caption{Procedure of building transmission backbone.}
        \label{Algorithm1}
        \begin{algorithmic}[1] %ÿÐÐÏÔʾÐкÅ
%            \Require $\alpha$ and $\$
%            \Ensure ÄæÐòÊý
        %%%ÊäÈëalpha and beta
        \State \emph{Vehicle} has packets to transmit
                \While{\emph{SV} is idle}
                     \State \textbf{Phase 1: Bootstrap phase}
                     \State Record the velocities and $\eta_{max}$ of surrounding vehicles
                     \State Calculate the $SI$ and obtain the minimum value to select $SV$ (\emph{e.g.}, vehicle A)
                     \State Inform Vehicle A that it has been selected as $SV$
                     \State Repeat until the backbone link is constructed
                     \State \textbf{Phase 2: Adaptive topology maintenance}
                     \State Calculate the $RI$ to identify the beacon period
                     \State $SV$ calculates the $SI$ values of neighbors and select the optimal replacement periodically
                     \State \textbf{Phase 3: Data transmission}
                     \State Source vehicle delivers packets to the nearest $SV$ to accomplish the multi-hop transmission.
                \EndWhile
                \While{\emph{SV} is not idle}
                     \State Source vehicle performs Phase 3
                \EndWhile
        \end{algorithmic}
    \end{algorithm}

\subsection{Bootstrap Phase}

All the vehicles in the network periodically broadcast beacon messages that contain the basic vehicle information including $ID$, \emph{location}, \emph{speed},
\emph{moving direction}, $\eta$ and $SV$. \emph{location} is defined as the Euclidian coordinates $(x,y)$. $\eta$ indicates the vehicle type. By
referring to \cite{Choo2004What}, we divide all vehicles into three types: \emph{$\eta$}=$1$, $2$, $3$ represent the compact, mid-size and large vehicles,
respectively. Specifically, by considering the similarity of vehicle size, the categories of small car and compact car in \cite{Choo2004What} are
involved in our compact vehicles category, and the large vehicle category includes all the full-sized vehicles in \cite{Choo2004What} while the mid-sized
vehicle category covers other categories in \cite{Choo2004What}. The indicator $SV$ shows that the vehicle belongs to $SV$s ($SV$=$1$) or
general vehicles ($SV$=$0$).

%This classification is obtained by simplifying the vehicles categories in \cite{Choo2004What}. {\color{red}{ In our study, the}} {\color{red}{ the categories of small car and compact car in \cite{Choo2004What} are involved in our compact vehicles category, and the large vehicle category includes all the full-sized vehicles in \cite{Choo2004What} while the mid-sized vehicle category covers other categories in \cite{Choo2004What}. Note that, the different classification of vehicle types does not change our conclusion that large vehicles incur more signal attenuation to small vehicles. Besides, a more detailed classification improves the accuracy of performance evaluation, but it also means a proportionate increase in work.}} $SV$ indicates that the vehicle belongs to $SV$s ($SV$=$1$) or general vehicles ($SV$=$0$).

The network is formed in the following steps. At the first beginning, $SV$s are not selected and the backbone link is not formed. The vehicle that has packets to transmit (called requesting vehicle) will evaluate the link stability to detect $SV$s among its neighbors based on the $SI$ (Stability Indication) evaluated by (\ref{stableindication}). Actually, the case that the requesting vehicle starts up the procedure of constructing backbone link has two basic conditions, including (1). it is in the bootstrap phase which means that there are no $SV$s on the road section or the nearest $SV$ is not within its communication range; (2). it does not detect the destination within its communication range. Upon identifying the $SV$, \emph{e.g.}, vehicle $A$, the requesting vehicle will send the \emph{INFORMATION} message to inform that it has been selected as a $SV$ and involved in the top tier link. Vehicle $A$ then sets its $SV$ indicator to $1$ and runs the same scheme to select the next top tier vehicle. By repeating above process, the backbone link will be constructed. Note, the requesting vehicle (or computing vehicle) does not consider whether their neighbors can communicate with the selected $SV$ or not. If the requesting vehicles on the road cannot detect $SV$s within the communication range, they will determine to select the new $SVs$ individually according to the beacon messages from neighbors. Otherwise, they connect to existing $SV$s directly.

\begin{equation}
\label{stableindication}
  SI_i=\alpha \left(\frac{|V_s-V_i|}{V_s}\right)+\beta\left(\frac{\eta_{\max}-\eta_{i}}{\eta_{\max}}\right).
\end{equation}

In (\ref{stableindication}), $\alpha$ and $\beta$ are weighting factors, determined by simulations. \emph{$V_s$} and \emph{$V_i$} are the velocities of the current computing vehicle and the neighbor \emph{i}, respectively, while the computing vehicle may be $SV$ or the first requesting vehicle. $\eta_{\max}$ is the maximum value of $\eta$. $\eta_{i}$ is the value of the neighbor \emph{i}. The first part of the $SI$ indicator, which is the determining factor of link duration, shows the relative speed between the current computing vehicle and its neighbors. It can be considered that two vehicles moving with small relative speed could connect for a relatively long link duration. The second part takes vehicle types into account and gives priority to large vehicles, which is an important impact factor of data rate. After computing $SI_i$, $i=1,2,...n,$ values of nearby vehicles, the vehicle that has minimum $SI$ will be selected as the $SV$.

To evaluate the link condition determined by (\ref{stableindication}) and the weighting factors, we first present the modeling of link duration by referring to \cite{togou2016scrp}. Let $D_{trans}$ denote the maximum transmission distance between communicating pairs while $D_{s,i}$ is the absolute value of the Euclidean distance. $V_s$ and $V_i$ are assumed as the instantaneous speeds of vehicle $s$ and $i$, respectively. We have
\begin{equation}
\label{LD1}
 {D_{trans}}-D_{s,i}=|V_s-V_i|LD_{s,i}.
\end{equation}
%where $LD$ is the estimating value of link duration according to current station.
%{\color{red}{delete}}
Then, we can obtain
\begin{equation}
\label{LD}
 LD_{s,i}=\frac{{D_{trans}-D_{s,i}}}{|V_s-V_i|}.
\end{equation}
It should be noted that the link condition is estimated periodically based on the initial $LD$ in each estimation period (beacon period).

We next formulate the evaluation as the following optimization model.
\begin{align}
\label{OptModel}
&\max\limits_{i=1,2...n}\,\, \hat{\Re}_{est}=LD_{_{s,i}}\xi_{s,i}\\
&s.t.\quad
\begin{cases}
\label{OptModelCase}
$C1$: LD_{_{s,i}}=\frac{{D_{trans}-D_{s,i}}}{|V_s-V_i|}, \\
$C2$: \xi_{s,i}=f{(P_{rTOT})}, \\
$C3$: \alpha+\beta=1, \\
\end{cases}
\end{align}
where $\xi_{s,i}$ denotes the data rate between $s$ and $i$, and $\hat{\Re}_{est}$ is the data volume calculated based on the link duration and data rate. Note that $\hat{\Re}_{est}$ is an estimation value in the estimation period  since both $LD$ and $\xi$ are instantaneous values at the initial instant. In (\ref{OptModelCase}), C2 denotes that $\xi_{s,i}$ is a function of $P_{rTOT}$, which means it can be obtained by the channel model in Section \ref{SectionIII}. Using the optimization model, the link condition can be evaluated at each estimation period. Furthermore, we conduct extensive simulations specifically for the determination of $\alpha$ and $\beta$ based on the model, as shown in Section \ref{SectionVI}.

\subsection{Adaptive Topology Maintenance}

As vehicles may randomly arrive and depart from the network, we develop an adaptive topology maintenance mechanism to combat the network dynamics. This
process is based on the exchange of beacon messages and implemented by transceiver $1$. Specifically, by the $SV$ indicators of beacons, $SV$s inform their neighbors the existence of the top tier link. Meanwhile, the route messages among $SV$s can also be updated periodically. In the maintenance mechanism, $SV$s compute \emph{SI} values for their neighbors periodically based on the received beacon messages. Once a new $SV$ is selected, its nearest $SV$ will be replaced and the route messages of the top tier are updated. As a matter of fact, this method can guarantee the advance supplement of $SV$s before the link breakage.

However, if there are no or less requesting vehicles on road, the frequent link maintenance becomes unnecessary. As such, the adaptive maintenance period based on the number of requesting vehicles is adopted. Specifically, we divide the beacon period into five classes based on the period range of $100$ms-$500$ms \cite{Haas2010Communication}. We define the \emph{Requesting Index} as the
ratio of the number of requesting vehicles ($\phi$)
to the total number of vehicles ($\Phi$) within the transmission range, i.e., $RI=\frac{\phi}{\Phi}$. According
to the received beacon messages, $SV$s get the values of $\phi$ and
$\Phi$ for their neighbors. The mapping table is shown in Table
\ref{Beaconperiod}. After determining the $RI$ range, $SV$s
broadcast the maintenance class to their neighbors. All vehicles, then, choose
the corresponding beacon period to broadcast beacon messages periodically.

\begin{table}[htbp]
 \caption{Setting of Beacon Period}
 \label{Beaconperiod}
 \centering
 \begin{tabular}{lclclclc}
  \toprule
 $RI$ Range  & Maintenance  Class & Beacon Period  \\
  \midrule
 $RI{\geq}0$ \&\& $RI{<}0.2$ & Class 1 & 500ms\\
  $RI{\geq}0.2$ \&\& $RI{<}0.4$ & Class 2 & 400ms\\
  $RI{\geq}0.4$ \&\& $RI{<}0.6$& Class 3  & 300ms\\
  $RI{\geq}0.6$ \&\& $RI{<}0.8$ & Class 4  & 200ms\\
  $RI{\geq}0.8$ \&\& $RI{<}1$ & Class 5  & 100ms\\
  \bottomrule
 \end{tabular}
\end{table}

\subsection{Data Transmissions}

%����֮һ�ǿ������������ɲ�ͬ�����е�organization ����ֻ��Ҫ��С���ڵڶ����ĸı�[reference, stable peers]
In the two-tier network framework, the intra-tier and inter-tier
communications can operate simultaneously without interference since we assume that each vehicle is equipped with two transceivers.

In the bottom tier, the one-hop data transmissions will be accomplished
directly if the source vehicle detects the destination within its transmission
range. Otherwise, the source
vehicle will find the nearest or optimal $SV$ and deliver packets to it. According the destination address, the $SV$ then selects the forwarding direction and transmit packets to the next $SV$. Upon
detecting the destination within the communication range, the $SV$
delivers the packets to it and the transmission is completed. In this process,
it is unnecessary for communication pairs to maintain the transmission route. In the
two-tier V2V communication architecture, all the maintenance mechanism needs
to do is maintain a backbone link so that the network overhead can be reduced.

\section{Analysis Model for Transmission Backbone}\label{SectionV}

After constructing the two-tier vehicular communication architecture, we are now ready to develop a comprehensive analytical framework based on queue theory to analyze the end-to-end performance of multi-hop data transmissions.

\subsection{Model of Queueing Network}

In this part, we assume the two-tier vehicular network as an open network of queues, and model each $SV$ as a $G/G/1$ queue. Based on the analysis framework, we then focus on end-to-end delay and achievable throughput in next subsection.

%%%%%%%%%%%%%%%%%%%%%%%%%%%%%%%%%%%%%%%%%%%%%%%%%%%%%%%%%%%%%%%%%%
%%%%%%%%%%%%%%%%%%%%%%%%%%%%%%%%%%%%%%%%%%%%%%%%%%%%%%%%%%%%%%%%%%
%¶ÔÀíÂÛ·ÖÎöÄ£ÐÍ×ö³õ²½µÄ¸ÅÊö
%¼ÙÉèÕû¸ö¹ý³ÌΪ¡£¡£¡£To describe the total communication process of two-tire network.
%¼ÙÉèÒµÎñµÄµ½´ïºÍÒµÎñµÄ·þÎñ¡£To assume the traffic arrival process and service process.
%¼ÙÉèÒµÎñµÄµ½´ïÊÇʲô£¬ÒµÎñµÄÊä³ö·ÖÊÇʲô¡£ To define that the arrival is consist of external arrival and internal arrival, while the spllitting is divided into external departure and internal departure
%¼ÙÉèÔÚÿ¸ö½ÚµãΪFCFS. To assume the FCFS displine
%È»ºóÎÒÃÇresort QNA£¬ÓÃÁ½¸ö²ÎÊýÀ´±íʾ. To resort the QNA: rate and variability.

\begin{figure*}[!tb]
\centering
\includegraphics[width=5in,height=1.4in]{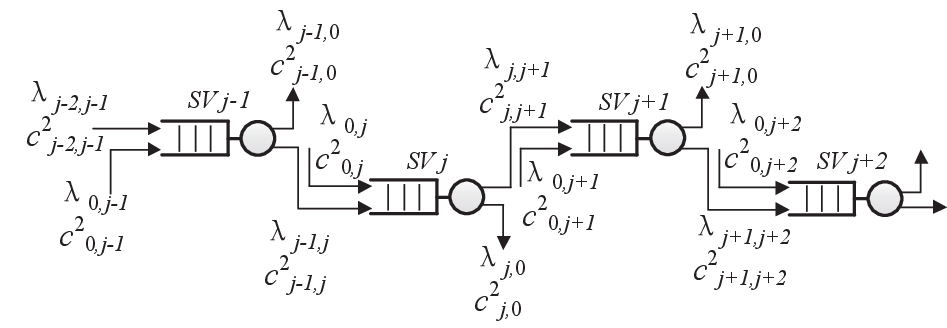}
\caption{Queueing network framework}%
\label{QueueNetwork}%
\end{figure*}

In the two-tier vehicular network, we consider that multi-hop packets, which come from $source$ $vehicle$ (general vehicle or $SV$) and eventually arrive at $destination$ $vehicle$ (general vehicle or $SV$), are forwarded by one or more $SV$s on the backbone link. In other words, each $SV$ may forward multiple traffic flows which are from multiple source vehicles to corresponding destination vehicles. Therefore, for each $G/G/1$ queue, the arrival process is modeled as a general distribution to allow arbitrary arrival patterns. Meanwhile, since $SV$s commonly act as routers for bottom tier vehicles, it is reasonable to assume that the service process at each $SV$ depends on the transmission rates to next-hop vehicles. Since the transmission rate is a function of distance, MAC layer contentions, and path loss, etc., we also model the service rate by a general distribution. In the backbone link, we consider the input of each $SV$ is consist of external arrival and internal arrival. The external arrival denotes the combination of several traffic flows generalized by different general vehicles, i.e., outside of the backbone link, while the internal arrival denotes traffic flows from the last hop $SV$. Similarly, the traffic flows of each $SV$ may be transmitted to the next $SV$ or general vehicles. For each $G/G/1$ queue, we assume the packets queued in each $SV$ node are in accordance with an FCFS (first-come-first-serve) discipline. Besides, the arrival rate is assumed as an independent and identically distributed (i.i.d.) random variable. The framework of the queueing network is shown in Fig. \ref{QueueNetwork}, in which each queue represents a $SV$ node. We next resort to the queueing network analyzer (QNA) to derive the expressions of our analysis framework \cite{whitt1983queueing}.

%%%%%%%%%%%%%%%%%%%%%%%%%%%%%%%%%%%%%%%%%%%%%%%%%%%%%%%%%%%%%%%%%%
%%%%%%%%%%%%%%%%%%%%%%%%%%%%%%%%%%%%%%%%%%%%%%%%%%%%%%%%%%%%%%%%%%
%½ÓÏÂÀ´¿ªÊ¼¾ßÌå·ÖÎö¹ý³Ì£º
%1ÎÒÃÇÖ÷Òª°Ñ·ÖÎö·ÖΪÁ½¸ö²ÎÊý£¬ËÙÂʺͷ½²îÀ´±íʾÁ½¸ö¹ý³Ì
%%%%%%%%%%%%%%%%%%%%%%%%%%%%%%%%%%%%%%%%%%%%%%%%%%%%%%%%%%%%%%%%%%
%%%%%%%%%%%%%%%%%%%%%%%%%%%%%%%%%%%%%%%%%%%%%%%%%%%%%%%%%%%%%%%%%%

We use two parameters to model the arrival rate and service rate, one to denote the mean and the other to describe the variability. This method is reasonable and feasible according to existing researches about queue theory \cite{whitt1983queueing}, \cite{bhat2015general}. In what follows, we give the detailed description about the derivation for arrival process and service process.

\subsubsection{External Arrival}

We begin with the derivation about the external arrival at a $SV$. For $SV_j$, let $\lambda_{0,j}$ and $c^2_{0,j}$ denote the mean and variance of external arrival rate, respectively. We have

\begin{equation}
\label{ex-arrival-rate}
\lambda_{0,j}=\sum_{k=1}^{\hat{r}_{j}}\hat{\lambda}_{k},
\end{equation}
where $\hat{\lambda}_{k}$ is the mean of traffic arrival rate from $k$-th general vehicle, and $\hat{r}_{j}$ is the number of general vehicles that are within the communication range of $SV_j$.

With $c_{a,j,k}^{2}$ denoting the variance of the traffic arrival rate from $k$-th general vehicle to $SV_j$, the variance $c^2_{0,j}$ is a function of the component variance $c_{a,j,k}^{2}$. We thus resort to a composite procedure based on convex combination and asymptotic method \cite{whitt1983queueing}, \cite{Albin1984Approximating}, \cite{whitt1982approximating}, where asymptotic method is a linear method to approximate the squared coefficients of variation of superposition, as described in Section 4.2 of \cite{whitt1982approximating}. Therefore, the variance of the external arrival rate at $SV_j$ can be obtained as

\begin{equation}
\label{ex-arrival-variability}
%c_{0,j}^{2}=\alpha_{j}\sum_{k=1}^{\hat{r}_{j}}\left(\frac{\hat{\lambda}_{k}}{\lambda_{0,j}}\right)c_{k}^{2}+1-\alpha_{j},
c_{0,j}^{2}=\alpha_{j}\sum_{k=1}^{\hat{r}_{j}}\left(\frac{\hat{\lambda}_{k}}{\lambda_{0,j}}\right)c_{a,j,k}^{2}+1-\alpha_{j},
\end{equation}
where $\alpha_{j}$ is the weight of convex combination and can be obtained as

\begin{equation}
\label{weight-alpha}
 \alpha_{j}=\left[1+4(1-\rho_j)^2(v_j-1)\right]^{-1},
\end{equation}
with
\begin{equation}
\label{weight-v}
   v_j=\left(\sum_{k}p_{k,j}^2\right)^{-1},
\end{equation}
where $k$ indicates the sequence of the surrounding vehicles of $SV_j$, including general vehicles and $SV_{j-1}$. Note that $\rho_j$ is the network utilization to indicate network states, which will be calculated in Subsection \ref{SubsectionV.B}. We denote by $p_{k,j}$ the proportion of arrivals came from $k$ to $j$\cite{whitt1983queueing}
\begin{equation}
\label{Proportion-P}
  % p_{k,j}=\lambda_{k,j}/\lambda_j,
   p_{k,j}=\frac{\lambda_{k,j}}{\lambda_j},
\end{equation}
where $\lambda_{k,j}$ characterizes the traffic flows from $SV_k$ to $SV_j$ , and $\lambda_j$ is the mean of overall arrival rates at $SV_j$ (including external arrival and internal arrival).

\subsubsection{Internal Arrival}

Having calculated the parameters of the external arrival, we now focus on the derivation of the internal arrival. In this part, let $\lambda_{j-1,j}$ denote the mean of internal arrival rate from $SV_{j-1}$ to $SV_j$. Apart from $\lambda_{j-1,j}$, we also consider the packets generalized by each $SV$ since it not only acts as a router but also may be a sender or receiver sometimes. From the backbone link perspective, we denote the generation or reception of each $SV$ by a constant factor $\delta$, which is the mean of packet generalized rates among $SV$s.

Hence, $\lambda_{j-1,j}$ can be obtained as

\begin{equation}
\label{in-arrival-rate}
\lambda_{j-1,j}=\left(\lambda_{j-1}+\delta\right)q_{j-1,j},
\end{equation}
where $q_{j-1,j}$ is the proportion of those packet flows completing service at $SV_{j-1}$ that go to $SV_j$.
% We assume
%\begin{equation}
%\label{q_j-1,j}
%%q_{j-1,j}=1-\hat{r}_{j-1}/(M-(j-1)+ \sum_{m=j}^{M}\hat{\lambda}_{m}),
%q_{j-1,j}=1-\frac{\hat{r}_{j-1}}{\left(M-(j-1)+ \sum_{m=j}^{M}\hat{r}_{m}\right)},
%\end{equation}
%where $M$ is the number of $SV$s on the backbone path.

We further resort to the splitting model in \cite{whitt1983queueing} to calculate the variance of the internal arrival rate $c^2_{j-1,j}$. We have

\begin{equation}
\label{in-arrival-variability}
c^2_{j-1,j}=c_{a,j-1}^{2}q_{j-1,j}+1-q_{j-1,j},
\end{equation}
where $c_{a,j-1}^{2}$\ is the variance of overall arrival rates in $SV_{j-1}$.

\subsubsection{Overall Arrival}
We now focus on the combination of the external arrival and internal arrival. Based on above expressions, we can obtain that the mean of overall arrival rates at $SV_j$ is

\begin{equation}
\label{total-arrival-rate}
\lambda_{j}=\lambda_{0,j}+\lambda_{j-1,j},
\end{equation}

and the variance is:%

\begin{equation}
\label{total-variance}
c_{a,j}^{2}=\alpha_{j}\left(p_{0,j}c_{0,j}^{2}+p_{j-1,j}c_{j-1,j}^{2}\right)+1-\alpha_{j}.%
\end{equation}

%%%%%%%%%%%%%%%%%%%%%%%%%%%%%%
%%%%%%%%%%%%%%%%%%%%%%%%%%%%%%
%ÓÃËæ»ú¹ý³ÌµÄ¾ùÖµºÍ·½²î¡£The mean and variance of stochastic process. ËùÒÔÊÇinter-arrival times and service times(ÌØÖ¸ÔÚij¸ö½ÚµãµÄ±äÁ¿Ê±¿ÉÒÔÓø´Êý)×÷ΪËæ»ú±äÁ¿¡£
%ºóÐøÓÃmean and variance of stochastic process to denote the random variables.
%%%%%%%%%%%%%%%%%%%%%%%%%%%%%%
%%%%%%%%%%%%%%%%%%%%%%%%%%%%%%

\subsubsection{Service Process}

\begin{figure}[!htb]
\centering
\includegraphics[width=2in,height=1.2in]{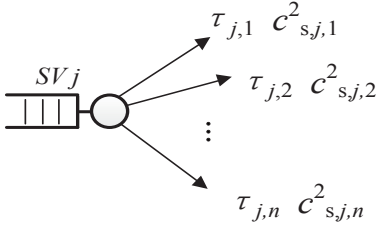}
\caption{Service process at queue j}
\label{serviceproess}
\end{figure}

Considering that the service process is determined by several stochastic factors such as random communicating distance, MAC layer contentions and path loss, we now concentrate on modeling the general distribution of service rate at $SV$. The service process of the queue $j$ is shown in Fig. \ref{serviceproess}, where $n$ is the number of next-hop vehicles including neighboring general vehicles and $SV_{j+1}$. Let the transmission rates from $SV_j$ to different neighboring vehicles, $\tau_{j,1}$, $\tau_{j,2}$, $\tau_{j,3}$, . . . $\tau_{j,n}$, be independent and identically distributed (i.i.d.), the mean of service rate at $SV_j$ can be calculated as

%\begin{equation}
%\label{servicerate}
%\tau_{j}=\sum_{k=1}^{\hat{r}_{j}}q_{j,k}\tau_{j,k}.
%\end{equation}
\begin{equation}
\label{servicerate}
\tau_{j}=\sum_{k=1}^{n}q_{j,k}\tau_{j,k}.
\end{equation}

Let $c^{2}_{s,j,k}$, $k=1,...n$, denote the variances of transmission rate ${\tau_{j,k}}$ from $SV_j$ to vehicle $k$, as shown in Fig. \ref{serviceproess}. If given $c_{s,j}^2$, which is the variance of service rate, we can obtain

\begin{equation}
\label{variance-k}
  c^{2}_{s,j,k}=q_{j,k}c_{s,j}^2+(1-q_{j,k}).
\end{equation}

Conversely, if we have $c_{s,j,k}^2$, $c_{s,j}^2$ can be calculated as

\begin{equation}
\label{servicevariability-k}
c^{2}_{s,j}=\frac{1}{q_{j,k}}\left(c_{s,j,k}^{2}-1\right)+1.
\end{equation}

It is reasonable to assume that the variance of the transmission rate between two adjacent $SV$s is easily obtained since there is real-time interaction between them. Therefore, given $c_{s,j,j+1}^2$, $c^{2}_{s,j}$ can be obtained
\begin{equation}
\label{servicevariability}
c^{2}_{s,j}=\frac{1}{q_{j,j+1}}\left(c_{s,j,j+1}^{2}-1\right)+1.
\end{equation}

\subsection{Performance Metrics}\label{SubsectionV.B}
\subsubsection{End-to-End Delay}

In this section, using our analysis framework, we first analyze the end-to-end delay of the proposed two-tier vehicular network architecture. The end-to-end delay for a packet transmitted from source to destination is consist of the waiting time at queues and the propagation time consumed on transmission path \cite{ramachandran2010queuing}.

Firstly, we resort to Little's Law to calculate the waiting time of each queue \cite{bhat2015general}. For classical $M/M/1$ queue, the waiting time is

\begin{equation}
\label{waitingtime}
  %E[WT]=\rho/\tau(1-\rho),
  E[WT]=\frac{\rho}{\tau(1-\rho)},
\end{equation}
where $\rho$ is the network utilization

\begin{equation}
\label{trafficuti}
  %\rho=\lambda/{\tau},
  \rho=\frac{\lambda}{\tau},
\end{equation}
which is assumed to satisfy $0\leq\rho<1$.

Therefore, for $j$-th $G/G/1$ queue in Fig. \ref{QueueNetwork}, the waiting time can be obtained by QNA \cite{whitt1983queueing}

\begin{equation}
\label{waitingtime-j}
  %E[WT_{j}]=\tau_{j}\rho_{j}(c_{a,j}^{2}+c_{s,j}^{2})g/[2(1-\rho_{j})],
  E[WT_{j}]=\frac{\rho_{j}(c_{a,j}^{2}+c_{s,j}^{2})g_j}{2\tau_{j}(1-\rho_{j})},
\end{equation}
where

\begin{equation}
\label{parameter-g}
g_j(\rho,c_{a,j}^{2},c_{s,j}^{2})=%
  \begin{cases}
    \exp[-\frac{2(1-\rho_{j})}{3\rho_{j}}\frac{(1-c_{a,j}^{2})^{2}}{c_{s,j}^{2}+c_{a,j}^{2}}], & c_{a,j}^{2}<1\\
        1, & c_{a,j}^{2}\geq1.
\end{cases}
\end{equation}

Considering the packets generalized by $SV_j$, the network utilization in the network is
\begin{equation}
\label{traffic-utilization}
  %\rho_{j}=\left(\lambda_j+\delta\right)/\tau_j.
  \rho_{j}=\frac{\lambda_j+\delta}{\tau_j}.
\end{equation}

Thus, the end-to-end delay between two communicating pairs is calculated as

\begin{equation}
\label{NetworkDelay}
  E[T]=E[PT]+E[WT],
\end{equation}
where $E[PT]$ is the propagation delay from source to destination.

Specifically, the end-to-end delay from vehicle $i$ to vehicle $j$ is

\begin{equation}
\begin{aligned}
E[T_{i,j}]&=\sum_{k=i}^{j}E[PT_{k.k+1}]+\sum_{m=i-1}^{j-1}E[WT_{m}] \\
          &=\sum_{k=i}^{j}\frac{1}{\tau_{k,k+1}}+\sum_{m=i-1}^{j-1}E[WT_{m}],
\end{aligned}
\end{equation}
where $m$ denotes the sequence of $SV$s on the given transmission path.

\subsubsection{Throughput}
We further analyze the achievable throughput of the selected backbone link. To obtain a comprehensive evaluation, we focus on the derivation of achievable throughput from the aspects of arrival rate and service rate of $SV$s. The average achievable throughput can be obtained as

\begin{equation}
%R_{Thr}=\min\Big\{\sum_{m}\lambda_{0,m},\min\{\tau_{k},k=1,2,3,..m\}\Big\}
%R_{Thr}= \sum_{j}^m\min\{\lambda_{0,j},{\tau_{j}}\}/m,
R_{Thr}= \sum_{j}^m\min\{(\lambda_{0,j}+\delta_j),{\tau_{j}}\},
\end{equation}
where $m$ is the number of $SV$s on the backbone link.
%If the network has infinite contention window, i.e., we do not consider the packet loss, then the max throughput can be obtained. The max throughput $R_{\max}$ is total external arrival rate since no packet loss occurs at each vehicle, which is denoted as:%
%
%\begin{equation}
%\label{MaxThroughput}
%R_{\max}=\sum_{m}\lambda_{0,m}.
%\end{equation}
%
%When the contention windows is finite, the packet loss can not be neglected. In this case, the throughput is the min value among the service rates at $SV$s on the selected path, which is denoted as:%
%
%\begin{equation}
%\label{MinThroughput}
%R_{\min}=\min\{\tau_{k},k=1,2,3,..m\},
%\end{equation}
%where $\tau_{k}$ is the service rate of $SV_k$.

%%%%%%%%%%%%%%%%%%%%%%%%%%%%%%%%%%%%%%%%%%%%%%%%%%%%%%%%%%%%%
%%%»áÒéʱµÄPerformance Evaluation,À©Õ¹Ê±È¥µô
%%%%%%%%%%%%%%%%%%%%%%%%%%%%%%%%%%%%%%%%%%%%%%%%%%%%%%%%%%%%%
\section{Performance evaluation}\label{SectionVI}

In this section, we evaluate the performance of our proposal using trace-driven simulations on Matlab.

%1¡¢ËùÓеijµÁ¾¶¼Óз¢ËÍÐÅÏ¢µÄÇëÇó
%2¡¢Ã¿¸ö³µµÄÐÅÏ¢²úÉúËÙÂÊÉèΪÏàͬ£¬µ«ÊÇc2²»Í¬
%3¡¢ÎÒÃÇͨ¹ý¶à´Î·ÂÕæÀ´µÃµ½Ã¿¸öSVµÄ·þÎñËÙÂʵÄmean and variance of service time
%4¡¢Unsaturation throughput£º¸ù¾Ýpacket generation rateÀ´ÅжÏ"It is well known that the network with 802.11 MAC
%5¡¢M is the maximum backoff stage. The retransmission limit is reached when the number of transmission failures of a packet reaches M+1;ÎÒÃÇÈ¡M=0£¬ËùÒÔ³öÏÖÒ»´Î´«Êäʧ°Ü¼´ÈÏΪ½«¸Ã°ü¶ªµô¡£
%scheme potentially exhibits an unstable behavior",from[throughput analysis cooperative mobile content distribution in vehicular network using symbol level network coding]

\subsection{Simulation Settings}

In all simulations, we set the network size to be $1000$ and the simulation time to be $90$ seconds. By referring to \cite{luan2013integrity}, we assume the maximum V2V communication range  as $300$ meters. Our mobility model is built on the vehicle trajectory from the real data set of US-$101$, as described in Section \ref{SectionII}. In the data set, there are average $31$ vehicles on the $640$ meters length of road section per second. Thus, we assume vehicles are placed on the road following Poisson distribution with the initial road traffic density is $\gamma=0.05$ veh/m \cite{ballerini1996probability}. For simplicity, we divide all the vehicles into three types: compact vehicles, mid-size vehicles and large vehicles, as illustrated in Section \ref{SectionIV}. According to the analysis results in Section \ref{SectionII}, we assume the speeds of type $1$ and type $2$ vehicles at each unit time follow normal distribution with the mean value uniformly distributes within [$20$, $40$] km/h and the standard deviation is $13$ km/h. For type $3$ vehicles, the speed at each unit time follows normal distribution with the mean value uniformly distributes within [$20$, $30$] km/h and the standard deviation is $9$ km/h. In this case, the average speed difference between large vehicles and other vehicles is within $15$ km/h, which is consistent with the results described in \cite{Ghods2012Effect}. Initially, we assume the percentages of large vehicles, mid-size vehicles and small vehicles are $5.4\%$, $34.7\%$ and $62.5\%$, respectively. This assumption is reasonable according to \cite{Choo2004What}. Even though the initial simulation settings are based on the practical vehicle trajectory, our proposal is easily to apply in different scenarios through adjusting several parameters, which is also presented in following performance comparison. The detailed parameters of physical layer and MAC layer are shown in TABLE \ref{Parameters}.

\begin{table}[!htb] %��'һ������environment��������λ����h,here ��
\caption{Setting of Simulation Parameters}
\centering
\label{Parameters}
\begin{tabular}{p{1.4cm}|p{1cm}|p{1.4cm}|p{1cm}} %������ÿһ�еĿ��ȣ�ǿ��ת����
\hline
\hline
Parameters & Value & Parameters & Value \\ % ��&���ָ���Ԫ�������� \\��ʾ������һ��

\hline %��һ�����ߣ������ľͶ���һ���ˣ�����һ����4 ������
       \emph{$P_t$} & 23 dBm & \emph{CW}       & 32\\
\hline
       \emph{$G_t$} & 1      & \emph{SlotTime} & 13${\mu}s$ \\
\hline
       \emph{$G_r$} & 1      & \emph{SIFS}     & 32${\mu}s$ \\
\hline
       \emph{$h_t$} & 0.1{m}  & \emph{DIFS}     & 32${\mu}s$ \\
\hline
       \emph{$h_r$}& 0.1{m}  & \emph{RTS}      & 53${\mu}s$ \\
\hline
       \emph{L}  & 1       & \emph{CTS}      & 37${\mu}s$ \\
\hline
\hline
\end{tabular}
\end{table}

\subsection{Evaluation of $SI$ Indicator}

\begin{figure}[!tb]
\centering
\includegraphics[width=3.1in]{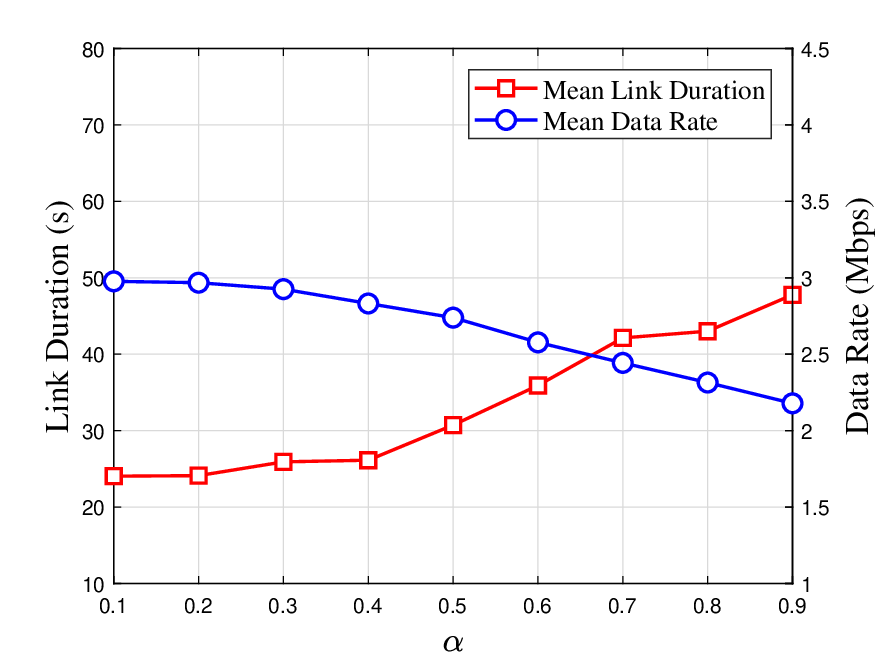}
\caption{Impact of $\alpha$ on link duration and data rate}
\label{alpha}
\end{figure}

%{\color{red}{We first conduct 1000 simulation runs in order to evaluate the $SI$ indicator and determine the optimal $\alpha$ and $\beta$, which is the preliminary of the construction of backbone link. Note, we do not require the obtained $\alpha$ and $\beta$ are always optimal for different mobility scenarios since the optimal values of $\alpha$ and $\beta$ in others scenarios can be redefined through \emph{our policy} shown in (\ref{stableindication})-(\ref{OptModelCase}). The variations of \emph{link duration} ($LD$) and data rate with $\alpha$ increasing are shown in Fig. \ref{alpha}. By increasing $\alpha$, we observe the decreased $LD$. This is because that the $SVs$ selected under high $\alpha$ have similar moving speed, accounting for long link duration. Conversely, these $SVs$ have small $\rho$ with the increase of $\alpha$. In this case, according to the analysis in Section II, there are low data rate between communicating pairs. To overcome the instability of $LD$ and data rate, we utilize the product of $LD$ and data rate to calculate the max data volume. Thus, the corresponding $\alpha$ is optimal, and $\beta=1-\alpha$. In the Fig. \ref{alpha}, the maximum data volume can be obtained in the point of $\alpha=0.7$.}}
In this part, we conduct $1000$ simulation runs in order to evaluate the $SI$ indicator and determine the optimal $\alpha$ and $\beta$, which are the preliminary of constructing the backbone link. Note, we do not require the obtained $\alpha$ and $\beta$ are always optimal since they may be different in other mobility scenarios where the optimal values can be redefined through our method. Specifically, we approximately choose the nine groups of values for $\alpha$ and $\beta$, as shown in Fig. \ref{alpha}. Firstly, a fixed target vehicle is selected randomly to act as the requesting vehicle in following simulations. For each pair of $\alpha$ and $\beta$, we select the first five $SV$s from the V2V link established by the requesting vehicle and calculate the average link duration and data rate. The corresponding results of link duration and data rate averaged by $1000$ simulations are then plotted in Fig.~\ref{alpha}. By substituting these results to the optimization model in (\ref{OptModel})-(\ref{OptModelCase}), the optimal $\alpha$ and $\beta$ can be obtained, i.e., $\alpha=0.7$ and $\beta=0.3$. In Fig. \ref{alpha}, by increasing $\alpha$, we observe the increased $LD$. This is because that the $SV$s selected under high $\alpha$ have similar moving speeds, accounting for a long link duration. However, the $SV$s with high $\alpha$ have small $\eta$, resulting in low data rate between communicating pairs according to the analysis in Section \ref{SectionII}.

%\subsection{Impact of $c_{a,j}^{2}$ and $c_{s,j}^{2}$}

\subsection{Impact of Variances on $G/G/1$}

\begin{figure}[!tb]
\centering
\includegraphics[width=3.3in]{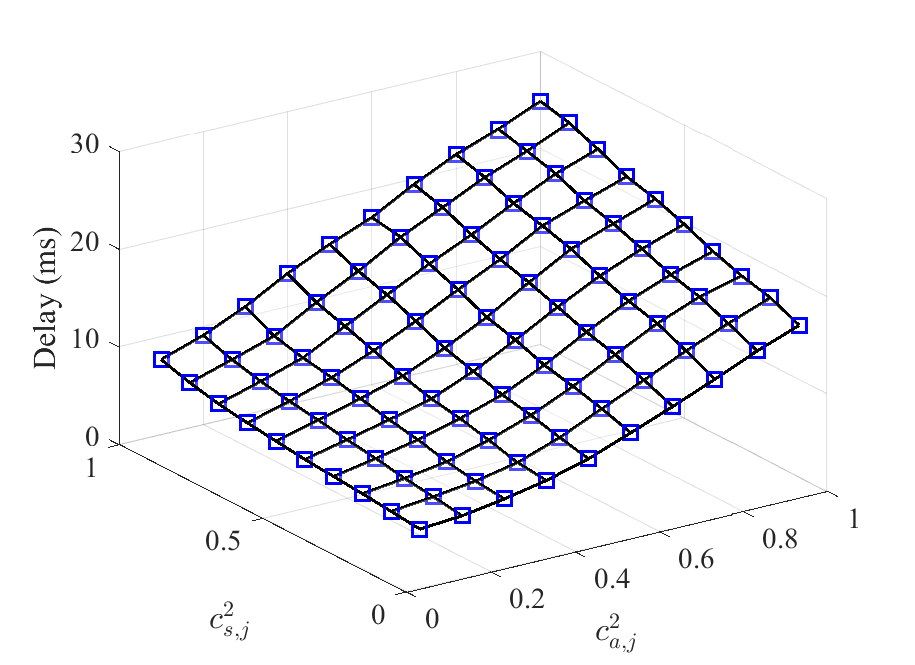}
\caption{Impact of variances on delay}
\label{VariancesifG}
\end{figure}

We further exploit the impact of the variances of queueing model on network performance. For simplicity, performance evaluation are performed by adjusting $c_{a,j}^{2}$ and $c^{2}_{s,j}$, which are the variances of overall arrival rate and service rate, respectively. Typically, $c_{a,j}^{2}=1$ denotes the arrival process is Poisson while $c^{2}_{s,j}=1$ shows the service-time distribution is exponential, as described in \cite{whitt1983queueing}. In this case, $G/G/1$ model is actually an $M/M/1$ model since $g_j(\rho,c_{a,j}^{2},c_{s,j}^{2})=1$ according to (\ref{parameter-g}). In this subsection, simulations are first conducted to construct the backbone link and obtain the first five $SV$s, where $\gamma=1/60$ veh/m and packet generation rate (PGR) is $20$ packets/s. Let all $SV$s have same $c_{a,j}^{2}$ and $c^{2}_{s,j}$. The delay is then calculated by the queueing model, as shown in Fig. \ref{VariancesifG}. As we can see, by increasing either $c_{a,j}^{2}$ or $c^{2}_{s,j}$ of queueing model, increased delay can be observed. When $c_{a,j}^{2}=1$ and $c^{2}_{s,j}=1$, $G/G/1$ model is transformed into $M/M/1$ model, and the delay attains the maximum value. This is because that $M/M/1$ simply considers the arrival process as the Poisson distribution. More generally, $G/G/1$ model considers all the arrival processes including Poisson or non-Poisson arrival process as renewal processes. Therefore, the $G/G/1$ model having a renewal arrival process independent of service times that are i.i.d. is more accurate and reasonable for the multi-hop data transmission architecture. By substituting the obtained $c_{s,j}^2$ into (\ref{variance-k}), $c^{2}_{s,j,k}$ can be calculated. Similarly, by substituting the obtained $c_{a,j}^2$ into (\ref{in-arrival-variability}), (\ref{total-variance}), and then into (\ref{ex-arrival-variability}), $c_{a,j,k}^{2}$ can be determined. Therefore, we choose $c^{2}_{s,j,k}=0.2$ and $c_{a,j,k}^{2}=0.2$ for the following performance evaluation.

%By combining the obtained results in this subsection and the analysis in section V, two values of $c^{2}_{s,j,k}=0.2$ and $c_{a,j,k}^{2}=0.2$ are chosen for following performance evaluation. {\color{red}{Specifically, by substituting the obtained $c_{s,j}^2$ into (\ref{variance-k}), $c^{2}_{s,j,k}$ then can be calculated. Similarly, by substituting the obtained $c_{a,j}^2$ into (\ref{in-arrival-variability}), (\ref{total-variance}), and then into (\ref{ex-arrival-variability}), $c_{a,j,k}^{2}$ can be determined.}}

\subsection{Performance Evaluation for Packet Delivery Ratio}

In this experiment, we demonstrate the effect of vehicle types on channel condition and the existence of stable channels. Based on the classification of vehicle types above, we extend the vehicle percentage into three classes, as shown in TABLE \ref{Percentage123}. Note that we do not require this is true in real scenario. We simply verify that our assumption that the impact of vehicle types on data transmissions is accurate.

\begin{table}[!b]
\renewcommand{\arraystretch}{1}
\newcommand{\tabincell}[2]{\begin{tabular}{@{}#1@{}}#2\end{tabular}}
\caption{Parameters for Three Classes of Vehicle Percentage}
\label{Percentage123}
\centering
\begin{tabular}{|c|c|c|c|c|}% |��ʾ��������
\hline%�±�����
\tabincell{c}{}&{Class 1} & \tabincell{c}{Class 2} & \tabincell{c}{Class 3}\\
\hline
  Large vehicles &\tabincell{c} 2.7\%       &\tabincell{c} 5.4\%         &\tabincell{c} 10.8\%\\
\hline
  Mid-size  Vehicles &\tabincell{c} 34.7\%   &\tabincell{c} 34.7\%         &\tabincell{c}34.7\%\\
\hline
  Small Vehicles     &\tabincell{c} 62.5\%  &\tabincell{c} 59.9\%         &\tabincell{c} 54.5\%\\
\hline
\end{tabular}
\end{table}

\begin{figure}[!tb]
\centering
\includegraphics[width=3.1in]{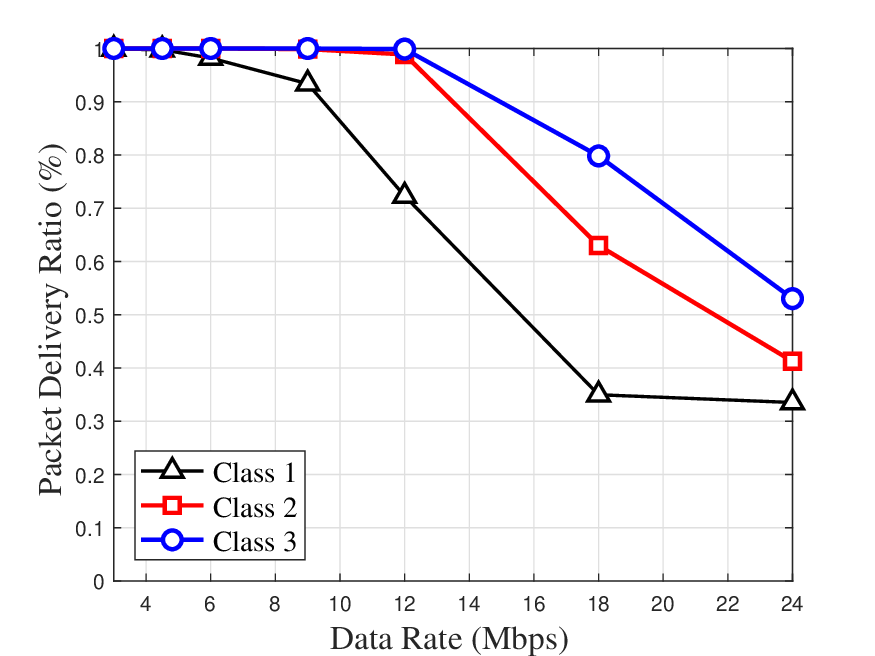}
\caption{Impact of classes of vehicle percentage on packet delivery ratio for various data rates}
\label{DeliveryRatio}
\end{figure}

Fig. \ref{DeliveryRatio} shows the results of the Packet Delivery Ratio (PDR) under different classes of vehicle percentages. In this experiment, we adopt the same communication pairs with previous simulations and the packet size is fixed to $800$ bytes. PDR is defined as the ratio of received packets to sent packets. Based on the minimum sensitivity thresholds as defined in DSRC standard (see TABLE \ref{SNR}) \cite{Astm2003Standard}, we obtain the PDR of the V2V link in our two-tier transmission architecture with different data rates. Typically, the optimal physical layer data rate is $6$ Mbps in VANET scenarios \cite{Jiang2008Optimal}. In this figure, when the data rate is higher than $6$ Mbps, the PDR presents a significant difference in different classes of vehicle percentages. Given the data rate of $12$ Mbps, we can see that the percentage of lost packets under \emph{Class 1} is about $29\%$ while this value under \emph{Class 3} is only $1\%$. From this experiment, we can demonstrate that vehicle type plays a significant role on the PDR of the two-tier communication architecture. Therefore, $SV$s involving large vehicles can enhance the stability of wireless channels. Moreover, the existence of stable channels can be verified.

%%%%%%%%%%%%%%%%%%%

\subsection{Performance Evaluation for Average E2ED}

We further carry out 1000 simulation runs to evaluate the average end-to-end delay (E2ED) for our two-tier vehicular network architecture. The average E2ED denotes the delay of a packet transmitted from source vehicle to destination. For simplicity, we report the delay of a packet forwarded by five $SV$s. Comparison is presented with a similar multi-hop transmission scheme called SCRP, as described in \cite{togou2016scrp}. We choose \cite{togou2016scrp} as it is closely related to our scheme in terms of the scheme of constructing backbone link. The highway or urban scenario is not the impact factor of network performance since our performance evaluation is conducted in a given road segment with multiple vehicle densities.

\begin{figure}
\centering
\subfigure[E2ED vs. PGR]{
\begin{minipage}[b]{0.5\textwidth}
\label{E2ED1}
\includegraphics[width=0.9\textwidth]{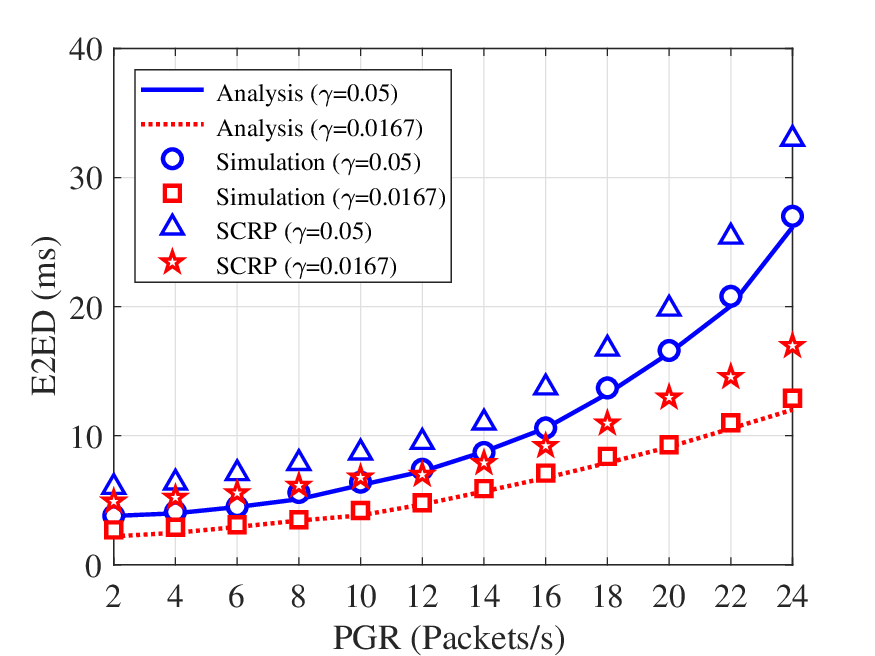}
\end{minipage}
}
\subfigure[E2ED vs. $\gamma$]{
\begin{minipage}[b]{0.5\textwidth}
\label{E2ED2}
\includegraphics[width=0.9\textwidth]{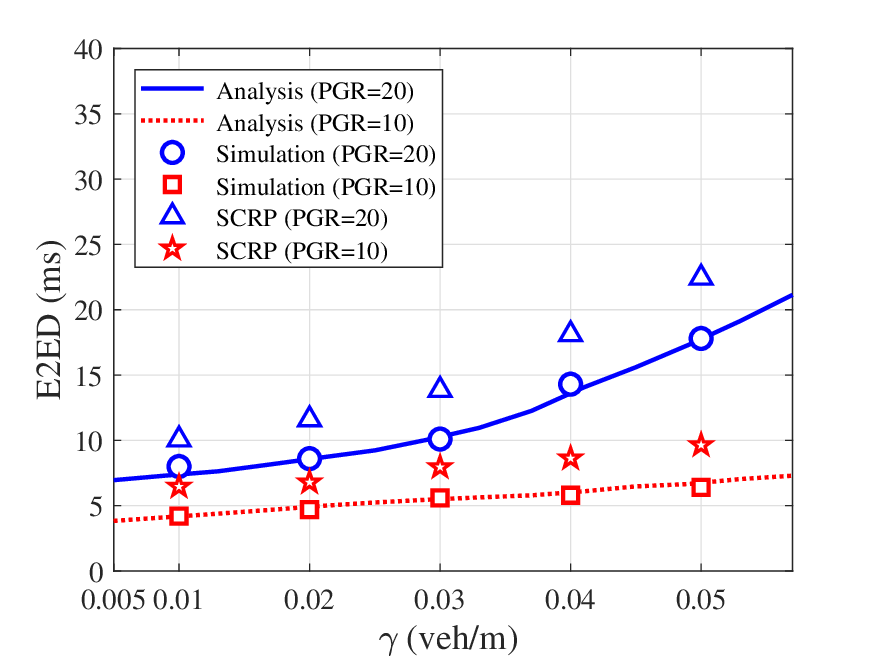}
\end{minipage}
}
% \caption{Relationship between E2ED, PGR and $\gamma$}
 \caption{Simulations comparison and analytic results for E2ED}
 \label{E2ED}
\end{figure}

Specifically, as shown in Fig. \ref{E2ED1}, with the PGR increasing, the results of the average E2ED obtained by both schemes increase and the increment also increases. The reason is that the increasing PGR results in more packet congestions. We can also see that our multi-hop transmission scheme always provides lower E2ED compared with SCRP under two scenarios with $\gamma=0.05$ vhe/m and $\gamma=0.0167$ vhe/m. This is because that our scheme combines channel conditions and inter-connection time to select $SV$s, which strengthen the data rate and V2V link duration. Fig. \ref{E2ED2} shows the relationship between the E2ED and the traffic density of vehicles ($\gamma$) with two different PGRs. As expected, with the $\gamma$ increasing, the average E2ED increases, especially in the scenario of
large PGR. As we can see, the large PGR results in faster increment of E2ED, which is consistent with the results in Fig. \ref{E2ED1}. Besides, the lower E2ED can also be
obtained by our scheme compared with SCRP. Both the figures in Fig. \ref{E2ED} show that the results obtained by queueing model based on $G/G/1$ match closely with simulation results. The accuracy of our theoretical analysis thus can be verified.

%\begin{figure}[!tb]
%\centering
%\includegraphics[width=3.8in,height=2.7in]{figure/PGRDelay.eps}
%\caption{E2ED vs. PGR}
%\label{E2ED1}
%\end{figure}
%
%\begin{figure}[!tb]
%\centering
%\includegraphics[width=3.8in,height=2.7in]{figure/VDDelay.eps}
%\caption{E2ED vs. $\gamma$}
%\label{E2ED2}
%\end{figure}

%\begin{figure}[!htb]
%  \centering
%  \subfigure[E2ED vs. PGR]
%  {
%    \label{E2ED1} %% label for first subfigure
%    \includegraphics[width=1.7in]{figure/PGRDelay.eps}
%  }
% \hspace{-4.5ex}
%  \subfigure[E2ED vs. $\gamma$]
%  {
%    \label{E2ED2} %% label for second subfigure
%    \includegraphics[width=1.7in]{figure/VDDelay.eps}
%  }
%  \caption{Relationship between E2ED, PGR and $\gamma$}
%  \label{E2ED} %% label for entire figure
%\end{figure}

\subsection{Performance Evaluation for Achievable Throughput}
%\begin{figure}
%\centering
%\subfigure[E2ED vs. PGR]{
%\begin{minipage}[b]{0.5\textwidth}
%\label{E2ED1}
%\includegraphics[width=1\textwidth]{figure/PGRThroughput.eps}
%\end{minipage}
%}
%\subfigure[E2ED vs. $\gamma$]{
%\begin{minipage}[b]{0.5\textwidth}
%\label{E2ED2}
%\includegraphics[width=1\textwidth]{figure/VDThroughput.eps}
%\end{minipage}
%}
% \caption{Relationship between E2ED, PGR and $\gamma$}
% \label{E2ED}
%\end{figure}
\begin{figure}
\centering
\subfigure[Achievable throughput vs. PGR]{
\begin{minipage}[b]{0.5\textwidth}
\label{Throughput:A}
\includegraphics[width=0.9\textwidth]{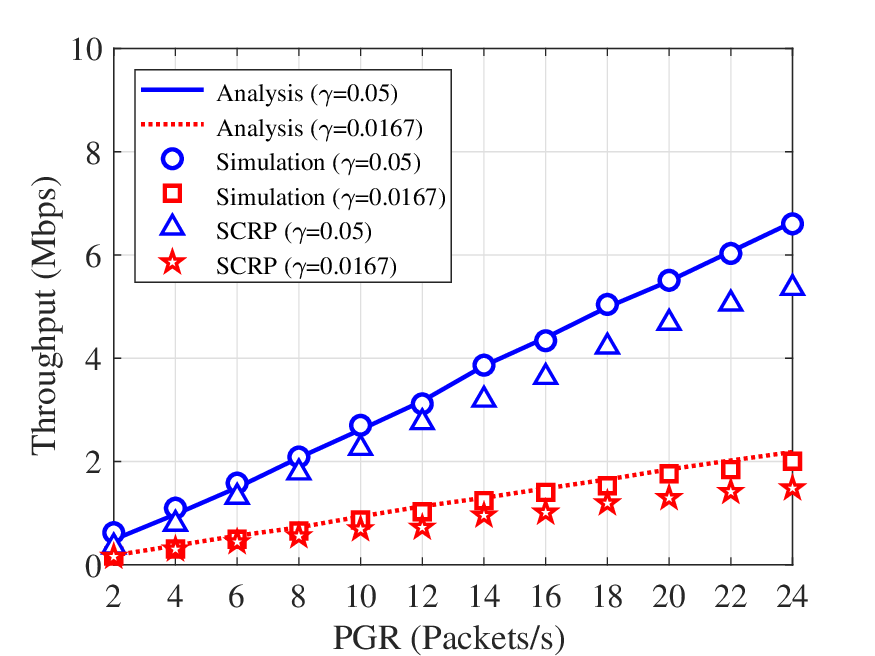}
\end{minipage}
}
\subfigure[Achievable throughput vs. $\gamma$]{
\begin{minipage}[b]{0.5\textwidth}
\label{Throughput:B}
\includegraphics[width=0.9\textwidth]{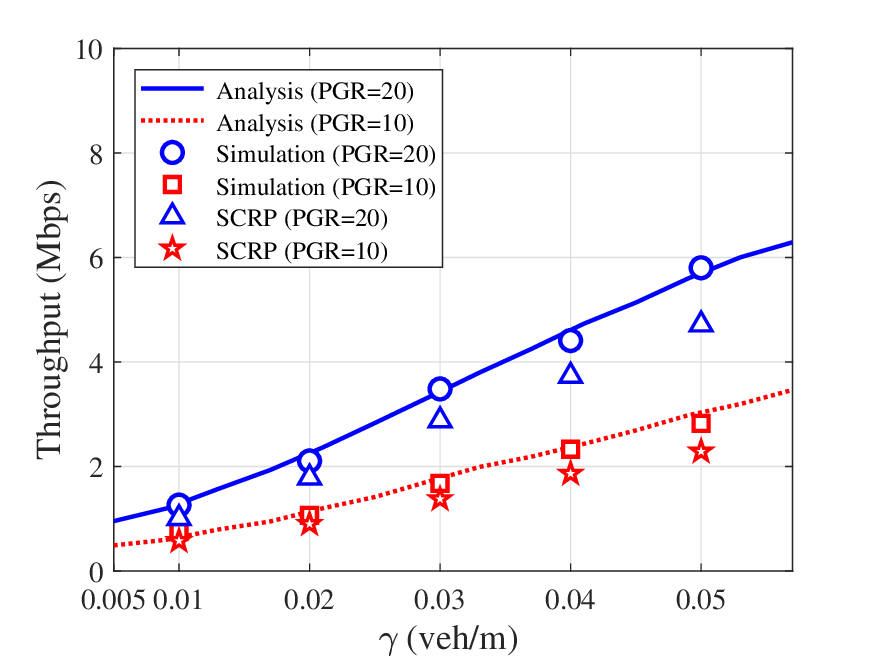}
\end{minipage}
}
 %\caption{Relationship between throughput, PGR and $\gamma$}
 \caption{Simulations comparison and analytic results for achievable throughput}
 \label{Throughput}
\end{figure}

In this experiment, we report the data volume forwarded through the five $SV$s in 90 seconds and compute the achievable throughput of the V2V communication link. Fig. \ref{Throughput} compares the achievable throughput between the two-tier multi-hop communication scheme and SCRP, and shows that the results of theoretical analysis match that of simulation.

As expected, the increasing PGR and $\gamma$ lead to the increase of throughput. Specifically, as shown in Fig. \ref{Throughput:A}, the throughput with $\gamma=0.05$ has a faster increment than that with $\gamma=0.0167$. This is because that the large $\gamma$ will increase the probability of selecting large vehicles as $SVs$ which will strength the channel stability and V2V connection time. Fig. \ref{Throughput:B} shows the relationship between the throughput and $\gamma$. It can be seen that the gap between our two-tier multi-hop communication scheme and SCRP increases with the increase of $\gamma$. Both Fig. \ref{Throughput:A} and Fig. \ref{Throughput:B} reveal the fact that when PGR and $\gamma$ are small, the proposed two-tier multi-hop communication scheme becomes less effective in improving the throughput compared with SCRP. This is due to the fact that in the scenario of low traffic density and low PGR, the probability that V2V links are obstructed by vehicles reduces and the probability of packet congestion is also reduced. $SVs$ thus have no obvious impact on the network performance. Meanwhile, the accuracy of our queueing model is reverified in Fig. \ref{Throughput} because of the close match between the results of the theoretical analysis and simulations.

To conclude, in \cite{togou2016scrp}, the stability factor is defined by considering  the relative location and relative speed between vehicles. Therefore, the difference between the stability in our scheme and SCRP is that we further analyze the impact of vehicles as obstacles on V2V communications based on the real-world vehicle traces, resulting in the increase of data rate and link duration.

\section{Related work}\label{SectionVII}

In highway infrastructure-less scenarios, since the deployment and maintenance cost of base stations or roadside units (RSUs) is huge, high-rate ubiquitous
infrastructure connections become very expensive and unpractical \cite{Zhuang2012On}. In this case, there are plenty of researches focus on the efficient V2V transmission methods including clustering schemes \cite{zhou2014chaincluster, Ucar2016Multihop}, cooperative vehicular networks \cite{chen2017throughput, Chen2016On, Lai2016SIRC}.

In terms of clustering schemes, Zhou \emph{et al.} \cite{zhou2014chaincluster} propose a linear clustering method for vehicles that have same requesting contents
from the infrastructure to extent the download volume of a single vehicle. Ucar \emph{et al.} \cite{Ucar2016Multihop} propose a relative mobility with respect to the neighboring vehicles based clustering method to improve network stability. Different from these two works where the cluster only occurs when vehicles have common interests or similar speeds, we consider the vehicle type as one of the metrics to select next hop and improve channel stability.

With the assistance of infrastructures, Chen \emph{et al.} \cite{chen2017throughput,Chen2016On} focus on the cooperative content download among multiple vehicles through V2I and V2V communications to facilitate the efficient data transmissions. Lai \emph{et al.} \cite{Lai2016SIRC} develop a secure incentive scheme based cooperative downloading strategy to ensure the fair and reliable data transmissions among cooperative vehicles. In above two works, the cooperative communication occurs when the infrastructure is within the communication range while our work focuses on enabling efficient multi-hop data transmissions  in infrastructure-less scenarios.

For urban environments, Togou \emph{et al.} \cite{togou2015novel, togou2016scrp} work on the distributed multi-hop transmission scheme in recent years. They take advantage of vehicles' mobility and the link lifetime estimation to construct a multi-hop route scheme. Therefore, the selected paths have high connectivity and low delivery delay. In our work, apart from the above two factors, we further consider the vehicle type to improve the channel conditions.

Apart from the highway scenario, there are also efficient data transmission schemes for other scenarios. Salkuyeh \emph{et al.} \cite{Salkuyeh2016An} propose an adaptive geographic routing scheme to discover several independent routes between source and destination vehicles for reliable multi-hop transmissions. Zhu \emph{et al.} \cite{Zhu2016Geographic} propose a greedy opportunity routing protocol for multilevel scenarios to alleviate the impact of viaducts, tunnels, and ramps, etc. on multi-hop data transmissions. These two works can not be applied in highway scenario since they do not consider the highly relative mobility among vehicles and topology maintenance overhead.

\section{Conclusion}\label{SectionVIII}

This paper focuses on ensuring high-rate multi-hop data transmissions over highway vehicular ad hoc networks. Specifically, we first present a comprehensive trajectory analysis for the existence and influence of $SV$s based on real-world highway scenarios. The results validate that $SV$s can be used to improve the stability of inter-connections and channel conditions. As such, we further propose a two-tier vehicular communication architecture, where the top tier is a backbone link constructed by connecting $SV$s and other vehicles are attached to it for multi-hop data transmissions. To the best of our knowledge, this is first study that utilizes vehicle type to strengthen the channel stability of V2V communications. Using simulations and examples, we verify that our proposal can outperform the SCRP scheme in terms of E2ED and throughput. For example, when PGR=$20$ packets/s and $\gamma=1/60$ ($0.0167$) veh/m, the comparison results show that the E2ED of two-tier communication architecture decreases about $28.24\%$ compared with SCRP while the throughput increases around $26.6\%$. We then develop a queueing model based analysis framework to evaluate the end-to-end performance of the layered multi-hop data transmission architecture. This, to the best of our knowledge, is also the first literature that applies the $G/G/1$ model to the multi-hop vehicular data transmissions, which provides a general guidance for the further application of queueing theory in VANET. Results show that our theoretical analysis can match the simulations well.

\end{document}